\documentclass[acmsmall]{acmart} 
\usepackage{tikz}
\usepackage{amsmath,amsfonts}
\usepackage{xurl}
\usepackage{balance}
\usepackage{listings}
\usepackage{calc}
\usepackage{bbding}
\usepackage{pifont}
\usepackage{wasysym}

\usepackage{xcolor}
\usepackage{graphicx}
\usepackage{caption,subcaption}
\usepackage{enumitem}
\usepackage{supertabular}
\usepackage{colortbl}
\usepackage{microtype}
\usepackage{fancyvrb}
\usepackage{multicol}
\usepackage{rotating}
\usepackage{enumitem}
\usepackage{xspace}
\usepackage{booktabs}
\usepackage{hyperref}

\usepackage[ruled,vlined,linesnumbered]{algorithm2e}
\usepackage{tikz}
\usepackage{mdframed}
\usepackage{multirow}
\usepackage{bigstrut}
\usepackage{mathtools}
\usepackage{mathpartir}
\usepackage{comment}
\usepackage{tcolorbox}
\usepackage{pifont}
\usepackage{verbatim}
\usepackage{boxedminipage2e}
\usepackage[justification=centering]{caption}
\usepackage{makecell}
\usepackage{pifont}
\usepackage{threeparttable}
\usepackage{hhline}
\usepackage{longtable}
\usepackage[
detect-weight,
exponent-product=\cdot,
locale = DE,
group-separator=.,
output-decimal-marker={.},
]{siunitx}

\usepackage{algorithmic}
\usepackage{textcomp}
\usepackage{array}
\usepackage{todonotes}
\usepackage{fontawesome5}
\usepackage{varwidth}
\usepackage{rotate}
\newcounter{rowno}
\newcommand{\ccmark}{\textcolor{green}{\ding{52}}}  
\newcommand{\xxmark}{\textcolor{red}{\ding{56}}}

\newcommand{\seagent}{\textsc{SE-Agent}\xspace}
\newcommand{\sweagent}{\textsc{SWE-Agent}\xspace}
\newcommand{\trae}{\textsc{Trae}\xspace}
\newcommand{\owl}{\textsc{OWL}\xspace}
\newcommand{\gptswarm}{\textsc{GPTswarm}\xspace}
\newcommand{\agentorchestra}{\textsc{AgentOrchestra}\xspace}
\newcommand{\openhands}{\textsc{OpenHands}\xspace}

\newcommand{\semagent}{\textsc{SemAgent}\xspace}
\newcommand{\experepair}{\textsc{ExpeRepair}\xspace}

\newcommand{\chatdev}{\textsc{ChatDev}\xspace}

\newcommand{\etal}{\textit{et al.}}

\newcommand{\codeff}[1]{\texttt{\small #1}}

\begin{document}

\title{A Comprehensive Empirical Evaluation of Agent Frameworks on Code-centric Software Engineering Tasks} 



\author{Zhuowen Yin}
\email{yinzhuowen@stumail.neu.edu.cn}
\orcid{0009-0005-4063-2509}
\authornote{Both authors contributed equally to this research.}
\affiliation{
  \institution{Institute of Al for Industries, Chinese Academy of Sciences}
  \streetaddress{168 Tianquan Road}
  \city{Nanjing}
  \state{Jiangsu}
  \country{China}
  \postcode{211135}
}
\affiliation{
  \institution{Northeastern University}
  \streetaddress{96 Jinzhai Road}
  \city{Shenyang}
  \state{Liaoning}
  \country{China}
  \postcode{110819}
}

\author{Cuifeng Gao}
\email{gcf20225162@mail.ustc.edu.cn}
\orcid{0000-0003-0672-1485}
\authornotemark[1]
\affiliation{
  \institution{University of Science and Technology of China}
  \streetaddress{96 Jinzhai Road}
  \city{Hefei}
  \state{Anhui}
  \country{China}
  \postcode{230026}
}

\author{Chunsong Fan}
\email{fanchunsong25@mails.ucas.ac.cn}
\orcid{0009-0009-4142-599X}
\affiliation{
  \institution{Institute of Al for Industries, Chinese Academy of Sciences}
  \streetaddress{168 Tianquan Road}
  \city{Nanjing}
  \state{Jiangsu}
  \country{China}
  \postcode{211135}
}

\author{Wenzhang Yang}
\authornote{corresponding author}
\email{wzhyang@iaii.ac.cn}
\orcid{0009-0006-7836-1246}
\affiliation{
  \institution{Institute of Al for Industries, Chinese Academy of Sciences}
  \streetaddress{168 Tianquan Road}
  \city{Nanjing}
  \state{Jiangsu}
  \country{China}
  \postcode{211135}
}

\author{Yinxing Xue}
\authornotemark[2]
\email{yxxue@ustc.edu.cn}
\orcid{0000-0002-2979-7151}
\affiliation{
  \institution{Institute of Al for Industries, Chinese Academy of Sciences}
  \streetaddress{168 Tianquan Road}
  \city{Nanjing}
  \state{Jiangsu}
  \country{China}
  \postcode{211135}
}

\author{Lijun Zhang}
\email{zhanglj@ios.ac.cn}
\orcid{0000-0002-3692-2088}
\affiliation{
  \institution{Institute of Al for Industries, Chinese Academy of Sciences}
  \streetaddress{168 Tianquan Road}
  \city{Nanjing}
  \state{Jiangsu}
  \country{China}
  \postcode{211135}
}


\begin{abstract}
The rapid advancement of large language models (LLMs) has led to the emergence of intelligent agents capable of reasoning, acting, and observing in dynamic environments. Unlike traditional automation tools or static LLM-based systems, agents combine decision-making and tool utilization to autonomously accomplish complex tasks, showing great potential in software engineering. However, existing studies largely focus on specific tasks or isolated aspects, providing an incomplete picture of agents’ practical capabilities.

To address this, we conduct a comprehensive empirical study evaluating seven general-purpose agent frameworks across three representative code-centric tasks: software development, vulnerability detection, and program repair. Each task is assessed using standard, widely adopted benchmarks to ensure objective and comparable evaluation. Agent performance is systematically analyzed from three complementary perspectives: effectiveness (task success), efficiency (execution process), and overhead (token consumption).

Our findings reveal distinct capability patterns and trade-offs among the evaluated frameworks. In terms of effectiveness, agents achieve moderate overall performance, with \openhands showing the most balanced results in software development, \gptswarm achieving the highest detection accuracy in vulnerability analysis, and only a subset of agents repairing about half of the issues in program repair. Regarding efficiency, \agentorchestra tends to exhibit the longest trajectories and the most correction attempts due to coordination overhead, whereas \openhands demonstrate stronger reflective reasoning abilities. For overhead, software development incurs the highest monetary cost, with \openhands consuming the most tokens but benefiting from input caching, while \gptswarm remains the most cost-efficient. Furthermore, we conduct an in-depth cross-analysis of the relationship between effectiveness and efficiency, exploring the underlying reasons behind their interplay. These findings guide both practical adoption and future research toward more autonomous and efficient software engineering agents.
\end{abstract}





\maketitle

\section{INTRODUCTION}

With the rapid advancement of large language models (LLMs), their level of intelligence has grown remarkably, driving a new wave of innovation in intelligent agents. Agents are not simply more complex LLMs; rather, the ability to use tools—a defining feature that distinguishes humans from animals—also distinguishes agents from plain LLMs~\cite{google-agent}. By leveraging external tools, agents can not only execute actions but also make decisions and manage workflows, thereby enabling them to handle more intricate tasks. The Reason–Act–Observe paradigm has emerged as a cutting-edge technique toward achieving artificial general intelligence (AGI)~\cite{gao2025survey}.

Recently, increasingly elaborate and sophisticated agent frameworks have demonstrated tremendous potential in the software engineering domain~\cite{agentsurvey2025}. They are being adopted throughout the software development life cycle~\cite{liu2024large}, particularly in code-centric software engineering tasks such as software development~\cite{chatdev}, vulnerability detection~\cite{llmsmartaudit}, and program repair~\cite{semagent}. Despite their widespread application, their practical capabilities in these tasks remain largely unclear to researchers and practitioners.

With the growing interest in agents, several empirical studies have begun to investigate the behavior and performance of agents from different perspectives. For instance, Gao \etal~\cite{gao2025single} compared the performance of multi-agent and single-agent systems across diverse domains beyond software engineering. In the context of automated program repair, Ceka \etal~\cite{ceka2025understanding} conducted an in-depth investigation into how agent decision-making workflows affect performance, while Meng \etal~\cite{meng2025empiricalstudyllmbasedagents} performed fine-grained experiments to assess the effectiveness of agents.
However, despite these contributions, existing studies remain insufficient for a comprehensive evaluation of agent capabilities in software engineering due to two main limitations:
(1) they focus on specific tasks, predominantly automated program repair; and
(2) they emphasize isolated aspects, such as traceability or effectiveness, rather than providing a holistic assessment.

To address these limitations, we conduct a comprehensive empirical study to systematically evaluate the practical performance of state-of-the-art agent frameworks across diverse code-centric software engineering tasks. Specifically, to ensure fair and consistent comparisons across multiple tasks, we focus on general-purpose agent frameworks that are designed to be adaptable across domains, rather than task-specific systems optimized for a single purpose. In total, seven representative frameworks are selected as the evaluation subjects. In addition, given the absence of a unified benchmark applicable to all code-centric tasks, we adopt widely used, high-quality benchmarks for each individual task: SRDD \cite{srdd} benchmark for software development task, LLM-SmartAudit benchmark \cite{llmsmartaudit} for vulnerability detection task, and SWE-bench Lite \cite{swebenchlite} benchmark for program repair task.

To systematically evaluate agent capabilities, we note that their workflows differ fundamentally from both LLM-based and traditional code-based tools. Unlike pre-trained systems, agents make autonomous and dynamic decisions during task execution. Hence, it is essential to assess not only what they achieve (functional effectiveness) but also how they achieve it (process efficiency). Moreover, the complex reasoning and decision-making inherent in agent execution often result in substantially higher token consumption compared to single-pass LLMs. Thus, we further investigate the overhead of task completion, including how token expenditure is distributed across reasoning and execution steps. Evaluating both the outcomes and the processes provides a holistic understanding of an agent framework’s true level of intelligence. Consequently, our study is guided by three complementary research questions addressing the \textit{effectiveness}, \textit{efficiency}, and \textit{overhead} of agent frameworks. 

Our results reveal notable distinctions across the three evaluation dimensions:

\textbf{Effectiveness (RQ1):}
Agents demonstrate moderate performance across all tasks. \openhands achieves the most balanced quality in software development, while \gptswarm leads in vulnerability detection with a 77
\% accuracy. In program repair, only a subset of agents successfully repair about half of the issues, leaving substantial room for improvement. We observe clear effectiveness trade-offs and underscore the ongoing challenges in fully automating complex, code-centric software engineering tasks.

\textbf{Efficiency (RQ2):} 
The execution efficiency varies across frameworks and tasks. \seagent(Iter-3) exhibits the longest steps across all experiments. \agentorchestra exhibits the longest correction sequences, whereas \openhands ranks second, demonstrating stable and convergent behavior.
Correction rates further reveal reflective adaptability. Specifically, both \agentorchestra achieve high correction rates across tasks, while \seagent(Iter-1) attains the better result in software development, suggesting that lightweight orchestration with focused reasoning can rival more complex designs. Overall, efficiency depends less on agent quantity than on reasoning depth, coordination strategy, and feedback integration.

\textbf{Overhead (RQ3):} Software development is the most expensive task in terms of monetary cost, with \agentorchestra being the most resource-intensive. Token consumption patterns vary significantly across frameworks. For instance, \openhands uses the most tokens overall but benefits from caching, while \gptswarm uniquely produces more output than input tokens. Breakdown analyses further reveal that planning and reflection stages dominate multi-agent workflows, in contrast to single-agent systems that concentrate their costs on execution and editing stages.

Additionally, we explore the relationship between effectiveness and efficiency from the perspectives of multi-agent and single-agent frameworks, analyzing the underlying factors driving these dynamics. Overall, this study provides actionable insights for both researchers and practitioners. For framework developers, our findings highlight opportunities to optimize agent architecture design, reasoning coordination, and token usage efficiency. For end users, the results offer practical guidance in selecting agent frameworks that best balance performance, efficiency, and cost according to their specific application needs.

\noindent
\vspace{1mm}
\textbf{Contributions.} Our contributions are summarized as follows:
\begin{itemize}
    \item We provide the first comprehensive comparison of seven general-purpose agent frameworks across three representative code-centric software engineering tasks.
    \item We systematically assess these agent frameworks across each task from three holistic perspectives: effectiveness, efficiency, and overhead.
    \item We present new empirical findings to guide practical adoption and future research. All experimental results are publicly available in our GitHub repository\footnote{\url{https://github.com/YCHYZW/Agents-for-Software-Engineering}}. 
\end{itemize}
\section{Background and Priliminaries}
\subsection{LLM-based Agents}
\label{subsec:bg-agents}
The term ``agent'' is used with varying meanings across different communities \cite{anthropic-agent}. In some contexts, it refers to highly autonomous systems that operate over long time spans and employ multiple tools to solve complex tasks. In others, the same term may describe a more prescriptive workflow wrapped around an LLM, offering limited autonomy and decision-making capability. Such inconsistent usage often blurs the line between general LLM automation and true agent architectures \cite{agentsurvey2025}, making it difficult to evaluate and compare existing work.

\vspace{1mm}
\noindent
\textbf{Definition.} To clearly establish the scope of this study, we adopt the definition of LLM-based agents from Google’s recent Agent White Paper~\cite{google-agent}. 

\begin{quotation}
\textit{“This combination of reasoning, logic, and access to external information that are all connected to a generative AI model invokes the concept of an agent.”}
\end{quotation}

According to the white paper, an agent is viewed as a system driven by a cognitive architecture that governs its behavior, interaction, and decision-making. This architecture consists of three essential components:
(1) \textit{Model}, which serves as the central decision-making engine, typically implemented using one or multiple instruction-following language models;
(2) \textit{Tools}, which empower the agent to interact with external data and services beyond pure text generation; and
(3) \textit{Orchestration}, which manages the iterative perception–reasoning–action loop, including planning, tool invocation, action execution, and reflection until the task goal is reached.
In this work, we therefore focus on agent frameworks that embody these components.

\vspace{1mm}
\noindent
\textbf{Categories.} Based on the scale of collaboration, agent systems can be divided into single-agent and multi-agent paradigms \cite{gao2025single}. Single-agent systems employ a centralized architecture in which one agent independently performs perception, reasoning, and decision-making. Their simplicity and low communication overhead make them suitable for applications such as personal assistants and basic chatbots. However, a single agent can easily become a bottleneck under high concurrency, and its generalization capability is fundamentally constrained by the capacity of the underlying LLM.
In contrast, multi-agent systems distribute work across multiple agents, each specializing in specific tasks to improve performance through division of labor, parallel execution, and dynamic coordination. Expanding the number of agents can further enhance system adaptability. Depending on interaction dynamics, multi-agent systems can be cooperative, competitive, or hybrid. Nevertheless, they introduce considerable coordination complexity and high resource consumption. In this paper, we investigate both single-agent and multi-agent systems to ensure a comprehensive and unbiased analysis of LLM-based agent frameworks.

\subsection{Code-centric Software Engineering Tasks}
The software engineering life-cycle typically comprises several sequential and iterative phases, including requirements analysis, design, implementation or development, testing and verification, deployment, and maintenance or evolution.
Each phase involves distinct objectives and artifacts, ranging from specifying system requirements and designing architecture to implementing, verifying, and maintaining the software.
In this study, we focus on the code-centric phases of the life-cycle, particularly those that involve direct manipulation of source code, such as \textit{code development}, \textit{vulnerability detection}, and \textit{program repair}.
These tasks serve as representative scenarios for evaluating the capability and adaptability of agent frameworks in software engineering contexts.

\vspace{1mm}
\noindent
\textbf{Software Development}
Software development is the systematic application of engineering principles, methods, and tools to the production and evolution of software \cite{basili1989software}. 
Its primary goal is to transform stakeholder requirements into operational, high-quality software. The process is conventionally structured by a software development life cycle that consists of interdependent phases including requirements elicitation and analysis, architectural design, implementation, verification and validation and deployment \cite{aggarwal2005software}. 
With the advent of generative AI, large language models have shown considerable potential to transform development workflows~\cite{jackson2025impact}. Agents aim to extend these capabilities further by supporting end-to-end coverage of the development life cycle. A high-quality solution generated by an agent should satisfy the stated requirements, incur low execution or generation cost, require minimal manual refinement and scale seamlessly to larger codebases~\cite{agentsurvey2025}.


\vspace{1mm}
\noindent
\textbf{Vulnerability Detection.} Vulnerability detection is a critical discipline within software security, focused on the systematic identification of security‑critical flaws in software systems~\cite{bessey2010afew}. Its primary objective is to proactively uncover and remediate weaknesses before they can be exploited by malicious actors, thereby safeguarding the confidentiality, integrity, and availability of the system. Recent advances in artificial intelligence have significantly accelerated research in this area~\cite{chakraborty2021deep}. Agent-based approaches further enhance detection by combining multi-agent co-inspection, integration of additional knowledge from tool executions, and traditional static analysis techniques~\cite{agentsurvey2025}, aiming to improve accuracy, diversify the types of detected vulnerabilities, and enhance overall performance.


\vspace{1mm}
\noindent
\textbf{Program Repair}
Program Repair, often referred to as Automated Program Repair (APR), is a subfield of Software Engineering that aims to automatically correct software faults~\cite{le2019automated}. Its primary objective is to mitigate the substantial manual effort, time, and cost associated with the debugging and maintenance phases. The typical APR pipeline involves fault localization, patch synthesis, and patch validation~\cite{liu2019tbar}. Learning-based APR treats program repair either as a translation task, in which buggy code is converted to correct code, or as a generation task, where the correct code is infilled within the buggy code context~\cite{zhang2023survey}. LLM-based agents generally follow an iterative paradigm, refining patches based on model or tool feedback until correctness is achieved, effectively balancing scope and accuracy~\cite{agentsurvey2025}.


\section{Methodology}
\label{sec:methodology}

\subsection{Research Questions}

\begin{figure}[t]
  \centering
  \vspace{3mm}
  \includegraphics[width=\linewidth]{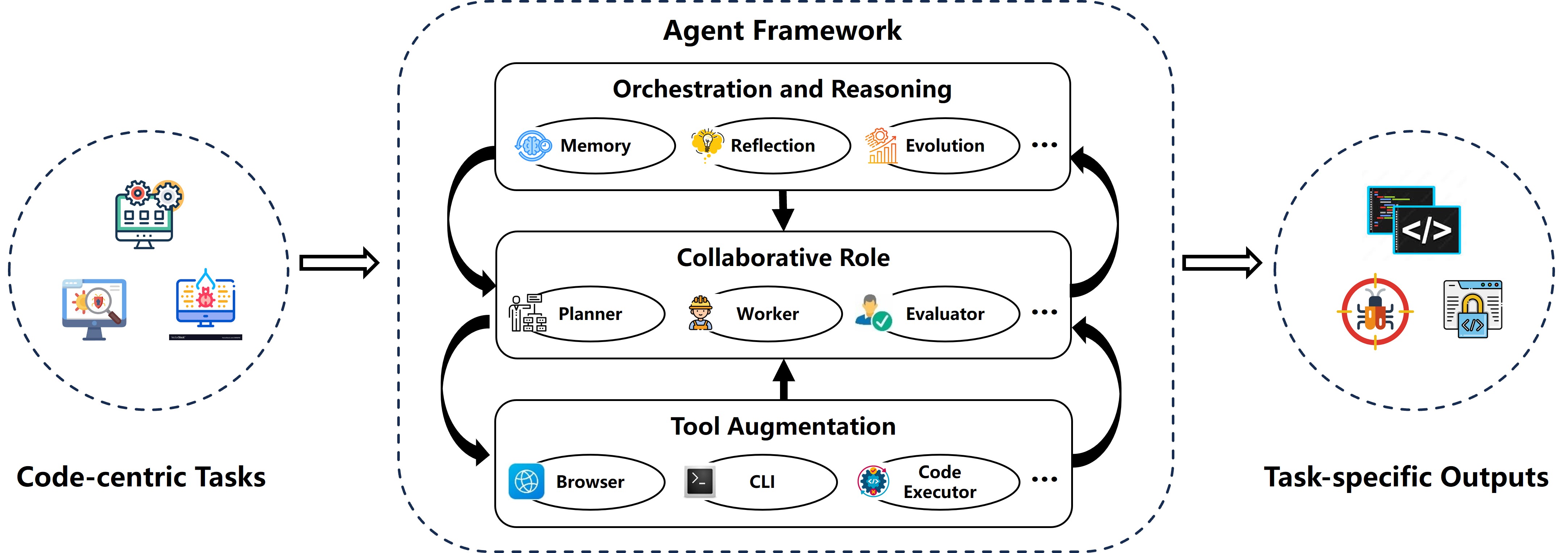}
  \caption{Agentic workflow paradigm.}
  \vspace{-3mm}
  \label{fig:workflow}
\end{figure}

To derive our research questions, we first abstract a generalized agentic workflow paradigm that characterizes how agent frameworks operate in code-centric software engineering tasks, as illustrated in \autoref{fig:workflow}. This paradigm adopts the cognitive architecture proposed in Google's Agent White Paper \cite{google-agent}, encompassing the model, tool, and orchestration components introduced in \S~\ref{subsec:bg-agents}. 
The agentic workflow comprises three conceptual layers that interact synergistically to enable adaptive, goal-driven task execution, as follows:

\vspace{1mm}
\noindent
\textbf{Orchestration and Reasoning.} This layer governs high-level decision-making and adaptive reasoning processes. It is responsible for dynamic task planning, self-reflection, and continuous evolution of strategies based on intermediate feedback. It typically incorporates memory mechanisms to retain contextual knowledge across iterations, enabling long-term consistency and informed reasoning. Reflection modules allow agents to analyze prior failures or inefficiencies, while evolution components iteratively refine prompts, goals, or workflows to enhance overall performance.

\vspace{1mm}
\noindent
\textbf{Collaborative Role} This layer defines specialized agent roles and their interactions within a cooperative framework. Common configurations include a planner that decomposes complex tasks into executable subtasks, workers that perform targeted actions or reasoning steps, and evaluators that assess the quality or correctness of intermediate outputs. This division of labor enables structured coordination and feedback loops among agents, thereby improving robustness and accountability in multi-agent collaboration.

\vspace{1mm}
\noindent
\textbf{Tool Augmentation} This layer extends the agent’s capabilities through external tools. These tools include browsers for retrieving up-to-date information from the web, command-line interfaces (CLI) for system-level operations, and code executors for testing or debugging generated code. Such augmentations bridge the gap between purely language-based reasoning and real-world execution, supporting more practical and context-aware problem solving.

These layers form an integrated ecosystem with rich interdependencies. The most distinctive feature of agent execution is its dynamic, adaptive intelligence, which differentiates agents from other approaches such as traditional code-based tools (e.g., static and dynamic analysis techniques) with fixed programs and static rules, or LLM-based tools that rely on manually engineered mechanisms (e.g., diversity chain-of-thought paradigms, fine-tuning, or retrieval-augmented generation), where processes and models are largely fixed.  


Consequently, a comprehensive and systematic evaluation of agent frameworks requires assessing three complementary aspects: the quality of their \textit{function}, the characteristics of their \textit{processes} underlying task execution, and the \textit{overhead} associated with their token consumption.
In terms of \textit{function}, we assess whether agents can effectively complete target tasks, focusing on final outcomes, as in traditional automation tools. 
In terms of \textit{process}, it is equally important to examine how agents achieve these goals, with particular emphasis on the efficiency and dynamics of their execution. This procedural aspect serves as a key indicator of the core intelligence level of the frameworks.
In terms of \textit{overhead}, it is not sufficient to measure only the total resources expended to complete a task; we must also analyze how resources are consumed across different stages of the reasoning process. Such a detailed examination provides deeper insights into the operational characteristics of agents and the significance of their execution strategies.

Based on this analysis, we formulate the following research questions:
\begin{itemize}
    \item \textbf{RQ1 [Effectiveness]:} How effectively do agent frameworks complete code-centric software engineering tasks?
    \item \textbf{RQ2 [Efficiency]:} How efficiently do agent frameworks utilize reasoning cycles and external tools to complete code-centric software engineering tasks?
    \item \textbf{RQ3 [Overhead]:} What is the token cost incurred by agent frameworks in completing code-centric software engineering tasks?

\end{itemize}


\subsection{Studied Agent Frameworks }
\label{subsec:agent}

\begin{table}[t]
\small
  \caption{Comparison of architectural components across the studied agent frameworks.}
  \label{tab:studied-agents}
  \setlength{\tabcolsep}{1pt}
  \resizebox{\textwidth}{!}{
  \begin{tabular}{cccccccccc}
    \toprule
    \textbf{Agents} & \textbf{Year} & \textbf{URL} & \makecell[c]{\textbf{Multi}\\\textbf{Agent}} & \makecell[c]{\textbf{Planner}} & \makecell[c]{\textbf{Context}\\\textbf{Builder}} & \makecell[c]{\textbf{Worker}} & \makecell[c]{\textbf{Evolution}}  & \makecell[c]{\textbf{Tool}\\\textbf{Selection}}& \makecell[c]{\textbf{Test-Time}\\\textbf{Scaling}}\\
    \midrule
    \agentorchestra \cite{agentorchestra} & 2025 
    & \url{https://github.com/SkyworkAI/DeepResearchAgent} 
    & \ccmark & \ccmark & \ccmark & \ccmark 
    & \xxmark & \ccmark & \xxmark\\
    \owl \cite{owl} & 2025 
    & \url{https://github.com/camel-ai/owl}
    & \ccmark & \ccmark & \ccmark & \ccmark 
    & \xxmark & \xxmark & \ccmark\\
    \seagent \cite{seagent} & 2025 
    & \url{https://github.com/JARVIS-Xs/SE-Agent}
    & \xxmark & \xxmark & \ccmark & \ccmark 
    & \ccmark & \xxmark & \ccmark\\
    \trae \cite{trae} & 2025 
    & \url{https://github.com/bytedance/trae-agent}
    & \xxmark & \xxmark & \xxmark & \ccmark 
    & \ccmark & \ccmark & \ccmark\\
    \gptswarm \cite{gptswarm} & 2024 
    & \url{https://github.com/metauto-ai/GPTSwarm} 
    & \xxmark & \xxmark & \ccmark & \ccmark 
    & \xxmark & \xxmark & \xxmark\\
    \openhands \cite{openhands} & 2024 
    &  \url{https://github.com/All-Hands-AI/OpenHands}
    & \xxmark & \xxmark & \ccmark & \ccmark 
    & \xxmark & \ccmark & \ccmark\\
    \sweagent \cite{sweagent}  & 2024 
    & \url{https://github.com/SWE-agent/SWE-agent}
    & \xxmark & \xxmark & \ccmark & \ccmark 
    & \xxmark & \ccmark & \xxmark\\
  \bottomrule
\end{tabular}
}
\end{table}

To comprehensively collect state-of-the-art agent frameworks that support code-centric software engineering tasks, we began with the survey by Liu \etal~\cite{liu2024large}, which provides an a comprehensive collection of research papers on LLM-based agents for software engineering. Note that this survey also includes LLM-based works that fall outside the scope of agents as defined in \S~\ref{subsec:bg-agents}. We thus carefully identified the relevant agent frameworks that meet our target criteria.

We then expanded this collection by searching both academic literature and industrial implementations available on the Internet. 
In line with the research goals of this study, which span diverse code-centric tasks, we prioritized general-purpose agent frameworks over ad hoc, task-specific systems, thereby enabling consistent and fair comparisons.
Therefore, only frameworks that satisfied the following inclusion criteria were considered in our study:
\begin{enumerate}[label=C\arabic*]
\item The agent framework can be applied to all three tasks, namely software development, vulnerability detection, and program repair.
This criterion excludes task-specific agents that are tailored to a single scenario, such as those exclusively designed for vulnerability repair or code generation.
\item The agent framework is publicly available.
This criterion excludes proprietary or closed-source frameworks that are not accessible for independent evaluation.
\end{enumerate}

In total, as shown in \autoref{tab:studied-agents}, we identified 7 that satisfy the inclusion criteria and summarizes the architectural components adopted by these frameworks.



\textbf{\agentorchestra} \cite{agentorchestra}  created by researchers at Skywork AI and Nanyang Technological University. It is a multi-agent framework that supports unified multimodal tool integration, flexible component composition, and efficient hierarchical collaboration. Through dynamic tool creation, intelligent retrieval, and automatic reuse.

\textbf{\owl} \cite{owl} developed by researchers from The University of Hong Kong and the CAMEL team, OWL is a hierarchical multi-agent framework. A Planner Agent decomposes tasks, a Coordinator Agent manages subtasks, and Worker Agents invoke domain-specific tools to execute work. Adding or modifying Worker Agents improves overall generalization during inference. Furthermore, \owl employs reinforcement learning with real-world feedback to optimize a domain-agnostic Planner Agent, thereby enhancing cross-domain generalization.

\textbf{\seagent} \cite{seagent} jointly developed by Tsinghua University and StepFun team. It is a self-evolving, multi-step reasoning agent. \seagent first constructs an initial trajectory pool by injecting five distinct reasoning strategies and controlled trajectory mutations. The agent then performs reflection and revision to identify issues, followed by cross-trajectory fusion via Crossover, Transfer, and Restructuring. Finally, it conducts multi-dimensional trajectory selection to produce robust solutions.

\textbf{\trae} \cite{trae} built by engineers at ByteDance for code engineering. It orchestrates a Code Agent that integrates multiple tools to iteratively generate candidate patches across prescribed steps. It then applies a two-stage pruning process—patch deduplication and regression testing—to reduce the search complexity for the optimal patch. To select candidates, a Selector Agent with repository-level understanding ranks solutions, and a voting mechanism determines the final patch.

\textbf{\gptswarm} \cite{gptswarm} proposed by researchers from KAUST and the Swiss AI Lab IDSIA. It models agents as computation graphs: nodes are functions that process multimodal data or query LLMs, edges represent interaction flows, and graphs can be recursively composed into larger hierarchies to capture inter-agent collaboration structures. \gptswarm introduces novel graph-level optimization operators: it refines node-level LLM prompts and adjusts graph connectivity to improve agent orchestration.

\textbf{\openhands} \cite{openhands} is an open-source platform co-developed by academic and industry contributors to provide a unified, powerful, and safe environment for building and evaluating general-purpose ``AI software developer'' agents. 1) It centers on an event-stream interaction mechanism that precisely logs each agent action and environment observation, forming complete task trajectories. 2) It bundles a ready-to-use Python tool library that encapsulates common but complex operations—file editing, multimodal parsing (e.g., PDFs, images), and code search—into simple function calls. 3) It pioneered a multi-agent delegation mechanism that allows a generalist agent (e.g., CodeAct Agent) to outsource specialized subtasks to expert agents (e.g., a Browsing Agent), enabling capability complementarity for complex composite tasks. As a result, \openhands supports diverse real-world scenarios, such as automated software development and maintenance, intelligent data analysis, efficient research assistance, and personal automation workflows.

\textbf{\sweagent} \cite{sweagent} developed by researchers at Princeton University. It is a single-agent framework tailored to software engineering. \sweagent augments an agent’s ability to create and modify files, navigate codebases, and run tests via an agent–computer interface (ACI). Its design mirrors human use of IDEs: (1) manual inspection of agent behaviors to identify failure modes and suggest improvements; and (2) grid search to select optimal ACI configurations. Together, these mechanisms strengthen the agent’s capacity to tackle complex software engineering tasks.

\subsection{Benchmark Suite}
\label{subsec:benchmark}
\begin{table}[t]
\small
  \caption{The benchmark suite for code-centric software engineering tasks.}
  \label{tab:benchmarks}
  \setlength{\tabcolsep}{3pt}
  \resizebox{\textwidth}{!}{
  \begin{tabular}{ccccc}
    \toprule
    \textbf{Task} & \textbf{Benchmark Dataset} & \textbf{Year} & \textbf{URL} & \textbf{Instances} \\
    \midrule
    Software Development & SRDD~\cite{srdd} & 2023 
    & \url{https://github.com/OpenBMB/ChatDev/tree/main/SRDD} 
    &  1,200 \\
    Vulnerability Detection & LLM-SmartAudit~\cite{llmsmartaudit} & 2025 
    & \url{https://github.com/LLMAudit/LLMSmartAuditTool} 
    & 115 \\
    Program Repair & SWEBench-Lite~\cite{swebenchlite} & 2024 
    & \url{https://www.swebench.com/lite.html} 
    & 300 \\
  \bottomrule
\end{tabular}
}
\end{table}

\label{subsec:dataset}

To ensure a rigorous and unbiased evaluation of the studied agent frameworks, we carefully curate a benchmark suite from publicly available software engineering benchmarks. Although agents are built on LLMs, LLM evaluation datasets and agent evaluation datasets differ fundamentally in their objectives, task granularity, and required interactions: the former measure intrinsic text understanding and generation abilities of LLMs, while the latter assess end-to-end task execution and autonomy in real workflows. Accordingly, agent evaluation typically incurs substantially higher costs than LLM evaluation. Therefore, we primarily consider the benchmarks used in prior software-engineering agent research, as catalogued in the survey by Liu \etal~\cite{liu2024large}. The selection was guided by the principles of broad community adoption, comprehensive task coverage, appropriate dataset scale, data reliability, and reproducibility to support consistent assessment and verification.

As summarized in \autoref{tab:benchmarks}, the final benchmark suite includes three benchmarks that span software development, vulnerability detection, and program repair tasks. In the following, we present the details of each benchmark and discuss the rationale behind its selection.


\vspace{1mm}
\textbf{SRDD~\cite{srdd} for Software Development.} To evaluate existing LLM-based agents for end-to-end software development, as reported by Liu~\etal~\cite{liu2024large}, many existing studies adopt classic code generation benchmarks such as HumanEval~\cite{humaneval} and MBPP~\cite{mbpp}. However, these benchmarks mainly involve simplified programming tasks on isolated pieces of code, which do not accurately represent practical development environments. To address this limitation, more advanced benchmarks have emerged, including SRDD~\cite{srdd} and ProjectDev~\cite{projectdev}, which support development tasks across multiple files and demonstrate increasing popularity in recent works.

Among these benchmarks, Software requirement description dataset (SRDD) has become the most frequently adopted one, since it contains the largest and most diverse collection of software development tasks with 1,200 curated prompts. Therefore, in this study, we select SRDD to evaluate agents on software development tasks. SRDD consists of 1,200 natural language prompts, organized into five major categories and 40 subcategories. The prompts were collected from ChatGPT 3.5 and thoroughly curated to ensure high task quality and relevance to realistic development scenarios.



\vspace{1mm}
\textbf{LLM-SmartAudit~\cite{llmsmartaudit} for Vulnerbaility Detection.}
Although many benchmarks have been used to evaluate LLM-based approaches for static bug detection, as reported by Liu \etal~\cite{liu2024large}, none are specifically curated for evaluating agent frameworks. In our investigation of recent works, LLM-SmartAudit is the only one we identified that applies agents for vulnerability detection tasks. Therefore, we select the dataset used by LLM-SmartAudit in this study to enable a fair and standardized comparison of different agent-based approaches under consistent evaluation conditions.

LLM-SmartAudit consists of three subset datasets: a common-vulnerability set, a real-world set, and a CVE set. Among them, only the common-vulnerability set and the CVE set provide standardized vulnerability type labels. In contrast, the real-world set is categorized based on their nature rather than their exact labels. Therefore, our evaluation focuses on the two well-labeled subsets, comprising 102 contracts from the common-vulnerability set and 13 contracts from the CVE set. In total, we analyze 115 vulnerability instances across these datasets.


\vspace{1mm}
\textbf{SWEBench-Lite~\cite{swebenchlite} for Program Repair.} 
For the task of program repair, many LLM-based approaches~\cite{liu2024large} have been evaluated on classical benchmarks such as Defects4J~\cite{defects4j} and QuixBugs~\cite{quixbugs}, which mainly target small-scale defects. In contrast, most agent-based works \cite{semagent, experepair, orcaloca} are consistently evaluated on the derivatives of the SWE-bench benchmark~\cite{swebench} (e.g., SWE-bench Lite~\cite{swebenchlite}, SWE-bench Lite-S~\cite{swebenchlites}, etc.), because it is grounded in real-world development scenarios by constructing tasks directly from GitHub issues and corresponding codebases, thereby ensuring practical relevance to authentic software maintenance workflows. 

SWE-bench~\cite{swebench} is a comprehensive benchmark, comprising over 2,000 real-world GitHub issues drawn from 12 widely-used Python repositories. Each task includes the original issue description, the full code repository, a predefined execution environment, and associated validation tests. However, evaluating the full SWE-bench benchmark incurs substantial computational costs and includes particularly challenging or problematic tasks~\cite{xia2024agentless}, which may lead to underestimation of the practical capability of LLM-based agents.

To address these limitations, researchers have manually curated high-quality tasks that balance difficulty, provide self-contained information, include informative issue descriptions, and offer sufficient evaluation tests, which constitutes SWE-bench Lite~\cite{swebenchlite}. Accordingly, we adopt SWE-bench Lite in this study to evaluate LLM-based agents on program repair tasks. This dataset comprises 300 carefully selected repair instances, covering all of the original 12 repositories while preserving the diversity and difficulty distribution of the full benchmark.



\subsection{Evaluation Framework}
\label{subsec:framework}

To ensure a fair and systematic evaluation across different agent frameworks, we design a unified evaluation framework that adopt tailored assessment metrics for each research question and standardizes the experimental setup for all agents.



\begin{table}[t]
\small
  \caption{Evaluation metrics for each research question.}
  \label{tab:metrics}
  \resizebox{\textwidth}{!}{
  \begin{tabular}{ccc}
    \toprule
    \textbf{Research Question} & \textbf{Task} & \textbf{Metrics}  \\
    \midrule
    \multirow{3}{*}{Effectiveness (RQ1)} 
    & Software Development &  $Quality = Completeness \times Executability \times Consistency$ \\
    & Vulnerability Detection & $Accuracy = TPs / (TPs + FPs)$ \\
    & Program Repair & $Repair\_Rate = (Correctly\_Fixed\_Issues) / (Total\_Issues)$ \\
    Efficiency (RQ2) & All Tasks & Average Trajectory Steps, Correction Attempts, and Correction Rate\\
    Overhead (RQ3) & All tasks & Monetary Cost, Token Usage \\
  \bottomrule
\end{tabular}
}
\end{table}

\vspace{1mm}
\textbf{Evaluation Metrics}
As shown in \autoref{tab:metrics}, we present the evaluation metrics corresponding to each research question. 
Effectiveness (RQ1) is measured using task-specific metrics aligned with the respective benchmarks. Specifically:
(1) On the SRDD benchmark, the evaluation script provided by ChatDev \cite{chatdev} leverages OpenAI’s text-embedding-ada-002 model to extract features from the project’s Python files, task descriptions, and requirement documents. This evaluation assesses the completeness, executability, and consistency of the generated code, which are combined into a final quality index.
(2) For the LLM-SmartAudit benchmark, agents generate vulnerability analysis reports that indicate the presence or absence of vulnerabilities and provide detailed classifications when applicable. These reports are compared against ground truth annotations to compute true positive (TP) and false positive (FP) rates.
(3) On the SWEBench-lite benchmark, we follow the official Docker-based evaluation protocol. The diff-format patch files generated by the seven agent frameworks are aggregated and standardized into a unified JSON format. Evaluation is performed via a command line interface (CLI), with repair rate as the primary metric. 

Efficiency (RQ2) is assessed by analyzing the agents’ execution processes. Specifically, we measure average trajectory steps, correction attempts, and correction rate. These metrics quantify how efficiently each agent framework completes the assigned tasks, capturing the dynamic decision-making behaviors intrinsic to intelligent agents.
Overhead (RQ3) measures the monetary cost incurred during task completion. We also record token consumption—including both input and output tokens—and analyze how these tokens are distributed across different stages of the agent workflow. This detailed breakdown offers insights into the cost-efficiency trade-offs among the evaluated frameworks.


\vspace{1mm}
\textbf{Experimental Setup.}
For the seven general-purpose agent frameworks, our experimental setup is primarily characterized by several key parameters, including, \textit{Step Limit}, \textit{Agent Quantity}, \textit{Tool Set}, \textit{Prompt}, \textit{Iteration} and \textit{Backend LLM}. The detailed configurations for each parameter are described as follows:
\begin{itemize}
    \item \textbf{Step Limit.} For all tasks and agent frameworks, the maximum step limit was set to 100 to ensure consistent termination conditions and to keep the monetary overhead within a controllable range.
    \item \textbf{Agent Quantity.} We adhere to the default configurations specified by each agent framework for executing code-related tasks. Notably, in \agentorchestra’s setup, all four agents are utilized for the software development task, whereas the BrowserUse agent is excluded from vulnerability detection and program repair tasks. This exclusion is due to the code-centric nature of these tasks, which rely primarily on direct source code interaction and static analysis within the repository, rendering browser capabilities unnecessary.
    \item \textbf{Tool Set} For all tasks and agents, the default tool sets provided by each framework were used without modification, ensuring an objective and faithful evaluation of their out-of-the-box capabilities.
    \item \textbf{Prompt.} The prompts are carefully aligned with the specific tasks and datasets to ensure relevance and consistency. To maintain comparability across frameworks, we employed prompt configurations previously validated on each respective benchmark. Specifically, for the software development task, we utilized prompts from \chatdev~\cite{chatdev}; for vulnerability detection, we adopted prompts from LLM-SmartAudit~\cite{llmsmartaudit}; and for the program repair task, we adapted prompts from \sweagent \cite{sweagent} with minor modifications to support all seven agent frameworks under evaluation.
    \item \textbf{Iteration.} Following \seagent's official example, three iterations are used with summary operators \codeff{null}, \codeff{alternative\_strategy} which generates distinct alternatives from recent failures, and \codeff{traj\_pool\_summary} which analyzes trajectory pool failures to identify blind spots and risks. \textit{We evaluate the performance across these three iterations and present the corresponding results in RQ1, RQ2, and RQ3 for comprehensive comparison.}
    \item \textbf{Backend LLM.} To ensure consistency and control costs, all agents utilize the DeepSeek-v3.1 \cite{deepseekai2024deepseekv3technicalreport} as their backend LLM, capitalizing on DeepSeek’s renowned innovation in training highly efficient models at low cost, thus offering an excellent price-performance ratio.
\end{itemize}

\section{Results and Analaysis}

In this section, we provide a comprehensive analysis of the empirical findings, focusing on the effectiveness (RQ1), efficiency (RQ2), and overhead (RQ3) of the seven general-purpose agent frameworks evaluated in this study.

\subsection{Effectiveness of Agent Frameworks (RQ1)}
\label{subsec:rq1}
To address RQ1, we leverage the benchmark suite introduced in \S~\ref{subsec:benchmark}, which covers three representative code-centric software engineering tasks: software development (SRDD), vulnerability detection (LLM-SmartAudit), and program repair (SWE-bench Lite). Across these tasks, we evaluate seven state-of-the-art agent frameworks, as presented in \S~\ref{subsec:agent}.

\begin{table}[t]
\small
  \caption{Comparison of software development effectiveness across different task classes and metrics on the SRDD benchmark.}
  \label{tab:sd-effectiveness}
  \setlength{\tabcolsep}{1pt}
  \resizebox{\textwidth}{!}{
  \begin{tabular}{lccccccccccccccccccc}
    \toprule
    \makecell[c]{\multirow{2}{*}{\textbf{Agent}}}
    & \multicolumn{3}{c}{\textbf{Game}} 
    & \multicolumn{3}{c}{\textbf{Education}} 
    & \multicolumn{3}{c}{\textbf{Work}} 
    & \multicolumn{3}{c}{\textbf{Life}} 
    & \multicolumn{3}{c}{\textbf{Creation}} 
    & \multicolumn{3}{c}{\textbf{Average}} 
    & \multirow{2}{*}{\textbf{Quality}}\\
    \cmidrule(lr){2-4} 
    \cmidrule(lr){5-7} 
    \cmidrule(lr){8-10} 
    \cmidrule(lr){11-13} 
    \cmidrule(lr){14-16} 
    \cmidrule(lr){17-19}
     & Comp. & Exec. & Cons. 
     & Comp. & Exec. & Cons. 
     & Comp. & Exec. & Cons. 
     & Comp. & Exec. & Cons. 
     & Comp. & Exec. & Cons. 
     & Comp. & Exec. & Cons.
     &
     \\
    \midrule
\agentorchestra & 
\cellcolor{darkgray!82}\textcolor{white}{0.82} & 
\cellcolor{darkgray!50}\textcolor{white}{0.50} & 
\cellcolor{darkgray!78}\textcolor{white}{0.78} & 
\cellcolor{darkgray!90}\textcolor{white}{0.90} & 
\cellcolor{darkgray!58}\textcolor{white}{0.58} & 
\cellcolor{darkgray!77}\textcolor{white}{0.77} & 
\cellcolor{darkgray!83}\textcolor{white}{0.83} & 
\cellcolor{darkgray!56}\textcolor{white}{0.56} & 
\cellcolor{darkgray!78}\textcolor{white}{0.78} & 
\cellcolor{darkgray!91}\textcolor{white}{0.91} & 
\cellcolor{darkgray!58}\textcolor{white}{0.58} & 
\cellcolor{darkgray!78}\textcolor{white}{0.78} & 
\cellcolor{darkgray!81}\textcolor{white}{0.81} & 
\cellcolor{darkgray!49}\textcolor{black}{0.49} & 
\cellcolor{darkgray!77}\textcolor{white}{0.77} &
\cellcolor{darkgray!86}\textcolor{white}{0.86} & 
\cellcolor{darkgray!55}\textcolor{white}{0.55} & 
\cellcolor{darkgray!78}\textcolor{white}{0.78} &
\cellcolor{darkgray!36}\textcolor{black}{0.36} \\

\owl & 
\cellcolor{darkgray!69}\textcolor{white}{0.69} & 
\cellcolor{darkgray!78}\textcolor{white}{0.78} & 
\cellcolor{darkgray!79}\textcolor{white}{0.79} & 
\cellcolor{darkgray!61}\textcolor{white}{0.61} & 
\cellcolor{darkgray!72}\textcolor{white}{0.72} & 
\cellcolor{darkgray!78}\textcolor{white}{0.78} & 
\cellcolor{darkgray!60}\textcolor{white}{0.60} & 
\cellcolor{darkgray!71}\textcolor{white}{0.71} & 
\cellcolor{darkgray!79}\textcolor{white}{0.79} & 
\cellcolor{darkgray!73}\textcolor{white}{0.73} & 
\cellcolor{darkgray!76}\textcolor{white}{0.76} & 
\cellcolor{darkgray!78}\textcolor{white}{0.78} & 
\cellcolor{darkgray!65}\textcolor{white}{0.65} & 
\cellcolor{darkgray!38}\textcolor{black}{0.38} & 
\cellcolor{darkgray!78}\textcolor{white}{0.78} &
\cellcolor{darkgray!66}\textcolor{white}{0.66} & 
\cellcolor{darkgray!70}\textcolor{white}{0.70} & 
\cellcolor{darkgray!78}\textcolor{white}{0.78} &
\cellcolor{darkgray!36}\textcolor{black}{0.36} \\

\seagent (Iter-1) & 
\cellcolor{darkgray!72}\textcolor{white}{0.72} & 
\cellcolor{darkgray!91}\textcolor{white}{0.91} & 
\cellcolor{darkgray!71}\textcolor{white}{0.71} & 
\cellcolor{darkgray!77}\textcolor{white}{0.77} & 
\cellcolor{darkgray!91}\textcolor{white}{0.91} & 
\cellcolor{darkgray!71}\textcolor{white}{0.71} & 
\cellcolor{darkgray!67}\textcolor{white}{0.67} & 
\cellcolor{darkgray!93}\textcolor{white}{0.93} & 
\cellcolor{darkgray!71}\textcolor{white}{0.71} & 
\cellcolor{darkgray!74}\textcolor{white}{0.74} & 
\cellcolor{darkgray!96}\textcolor{white}{0.96} & 
\cellcolor{darkgray!72}\textcolor{white}{0.72} & 
\cellcolor{darkgray!64}\textcolor{white}{0.64} & 
\cellcolor{darkgray!65}\textcolor{white}{0.65} & 
\cellcolor{darkgray!71}\textcolor{white}{0.71} &
\cellcolor{darkgray!71}\textcolor{white}{0.71} & 
\cellcolor{darkgray!90}\textcolor{white}{0.90} & 
\cellcolor{darkgray!71}\textcolor{white}{0.71} &
\cellcolor{darkgray!46}\textcolor{black}{0.46} \\

\seagent (Iter-2) & 
\cellcolor{darkgray!71}\textcolor{white}{0.71} & 
\cellcolor{darkgray!43}\textcolor{white}{0.43} & 
\cellcolor{darkgray!78}\textcolor{white}{0.78} & 
\cellcolor{darkgray!78}\textcolor{white}{0.78} & 
\cellcolor{darkgray!34}\textcolor{black}{0.34} & 
\cellcolor{darkgray!78}\textcolor{white}{0.78} & 
\cellcolor{darkgray!73}\textcolor{white}{0.73} & 
\cellcolor{darkgray!29}\textcolor{black}{0.29} & 
\cellcolor{darkgray!78}\textcolor{white}{0.78} & 
\cellcolor{darkgray!76}\textcolor{white}{0.76} & 
\cellcolor{darkgray!35}\textcolor{black}{0.35} & 
\cellcolor{darkgray!78}\textcolor{white}{0.78} & 
\cellcolor{darkgray!81}\textcolor{white}{0.81} & 
\cellcolor{darkgray!46}\textcolor{black}{0.46} & 
\cellcolor{darkgray!79}\textcolor{white}{0.79} &
\cellcolor{darkgray!75}\textcolor{white}{0.75} & 
\cellcolor{darkgray!37}\textcolor{black}{0.37} & 
\cellcolor{darkgray!78}\textcolor{white}{0.78} &
\cellcolor{darkgray!22}\textcolor{black}{0.22} \\

\seagent (Iter-3) & 
\cellcolor{darkgray!73}\textcolor{white}{0.73} & 
\cellcolor{darkgray!23}\textcolor{black}{0.23} & 
\cellcolor{darkgray!79}\textcolor{white}{0.79} & 
\cellcolor{darkgray!79}\textcolor{white}{0.79} & 
\cellcolor{darkgray!25}\textcolor{black}{0.25} & 
\cellcolor{darkgray!79}\textcolor{white}{0.79} & 
\cellcolor{darkgray!71}\textcolor{white}{0.71} & 
\cellcolor{darkgray!18}\textcolor{black}{0.18} & 
\cellcolor{darkgray!79}\textcolor{white}{0.79} & 
\cellcolor{darkgray!80}\textcolor{white}{0.80} & 
\cellcolor{darkgray!29}\textcolor{black}{0.29} & 
\cellcolor{darkgray!79}\textcolor{white}{0.79} & 
\cellcolor{darkgray!76}\textcolor{white}{0.76} & 
\cellcolor{darkgray!34}\textcolor{black}{0.34} & 
\cellcolor{darkgray!79}\textcolor{white}{0.79} &
\cellcolor{darkgray!76}\textcolor{white}{0.76} & 
\cellcolor{darkgray!26}\textcolor{black}{0.26} & 
\cellcolor{darkgray!79}\textcolor{white}{0.79} &
\cellcolor{darkgray!15}\textcolor{black}{0.15} \\

\trae & 
\cellcolor{darkgray!48}\textcolor{black}{0.48} & 
\cellcolor{darkgray!96}\textcolor{white}{0.96} & 
\cellcolor{darkgray!81}\textcolor{white}{0.81} & 
\cellcolor{darkgray!50}\textcolor{white}{0.50} & 
\cellcolor{darkgray!94}\textcolor{white}{0.94} & 
\cellcolor{darkgray!81}\textcolor{white}{0.81} & 
\cellcolor{darkgray!51}\textcolor{white}{0.51} & 
\cellcolor{darkgray!91}\textcolor{white}{0.91} & 
\cellcolor{darkgray!81}\textcolor{white}{0.81} & 
\cellcolor{darkgray!64}\textcolor{white}{0.64} & 
\cellcolor{darkgray!93}\textcolor{white}{0.93} & 
\cellcolor{darkgray!81}\textcolor{white}{0.81} & 
\cellcolor{darkgray!43}\textcolor{black}{0.43} & 
\cellcolor{darkgray!90}\textcolor{white}{0.90} & 
\cellcolor{darkgray!81}\textcolor{white}{0.81} &
\cellcolor{darkgray!53}\textcolor{white}{0.53} & 
\cellcolor{darkgray!93}\textcolor{white}{0.93} & 
\cellcolor{darkgray!81}\textcolor{white}{0.81} &
\cellcolor{darkgray!40}\textcolor{black}{0.40} \\

\gptswarm & 
\cellcolor{darkgray!76}\textcolor{white}{0.76} & 
\cellcolor{darkgray!64}\textcolor{white}{0.64} & 
\cellcolor{darkgray!85}\textcolor{white}{0.85} & 
\cellcolor{darkgray!69}\textcolor{white}{0.69} & 
\cellcolor{darkgray!81}\textcolor{white}{0.81} & 
\cellcolor{darkgray!85}\textcolor{white}{0.85} & 
\cellcolor{darkgray!71}\textcolor{white}{0.71} & 
\cellcolor{darkgray!77}\textcolor{white}{0.77} & 
\cellcolor{darkgray!85}\textcolor{white}{0.85} & 
\cellcolor{darkgray!80}\textcolor{white}{0.80} & 
\cellcolor{darkgray!58}\textcolor{white}{0.58} & 
\cellcolor{darkgray!85}\textcolor{white}{0.85} & 
\cellcolor{darkgray!77}\textcolor{white}{0.77} & 
\cellcolor{darkgray!81}\textcolor{white}{0.81} & 
\cellcolor{darkgray!86}\textcolor{white}{0.86} &
\cellcolor{darkgray!75}\textcolor{white}{0.75} & 
\cellcolor{darkgray!70}\textcolor{white}{0.70} & 
\cellcolor{darkgray!85}\textcolor{white}{0.85} &
\cellcolor{darkgray!45}\textcolor{black}{0.45} \\

\openhands & 
\cellcolor{darkgray!62}\textcolor{white}{0.62} & 
\cellcolor{darkgray!100}\textcolor{white}{1.00} & 
\cellcolor{darkgray!80}\textcolor{white}{0.80} & 
\cellcolor{darkgray!59}\textcolor{white}{0.59} & 
\cellcolor{darkgray!100}\textcolor{white}{1.00} & 
\cellcolor{darkgray!79}\textcolor{white}{0.79} & 
\cellcolor{darkgray!49}\textcolor{black}{0.49} & 
\cellcolor{darkgray!100}\textcolor{white}{1.00} & 
\cellcolor{darkgray!78}\textcolor{white}{0.78} & 
\cellcolor{darkgray!68}\textcolor{white}{0.68} & 
\cellcolor{darkgray!100}\textcolor{white}{1.00} & 
\cellcolor{darkgray!80}\textcolor{white}{0.80} & 
\cellcolor{darkgray!59}\textcolor{white}{0.59} & 
\cellcolor{darkgray!100}\textcolor{white}{1.00} & 
\cellcolor{darkgray!79}\textcolor{white}{0.79} &
\cellcolor{darkgray!60}\textcolor{white}{0.60} & 
\cellcolor{darkgray!100}\textcolor{white}{1.00} & 
\cellcolor{darkgray!79}\textcolor{white}{0.79} &
\cellcolor{darkgray!47}\textcolor{black}{0.47} \\

\sweagent & 
\cellcolor{darkgray!53}\textcolor{white}{0.53} & 
\cellcolor{darkgray!84}\textcolor{white}{0.84} & 
\cellcolor{darkgray!78}\textcolor{white}{0.78} & 
\cellcolor{darkgray!64}\textcolor{white}{0.64} & 
\cellcolor{darkgray!84}\textcolor{white}{0.84} & 
\cellcolor{darkgray!77}\textcolor{white}{0.77} & 
\cellcolor{darkgray!58}\textcolor{white}{0.58} & 
\cellcolor{darkgray!90}\textcolor{white}{0.90} & 
\cellcolor{darkgray!78}\textcolor{white}{0.78} & 
\cellcolor{darkgray!61}\textcolor{white}{0.61} & 
\cellcolor{darkgray!94}\textcolor{white}{0.94} & 
\cellcolor{darkgray!80}\textcolor{white}{0.80} & 
\cellcolor{darkgray!47}\textcolor{black}{0.47} & 
\cellcolor{darkgray!85}\textcolor{white}{0.85} & 
\cellcolor{darkgray!79}\textcolor{white}{0.79} &
\cellcolor{darkgray!57}\textcolor{white}{0.57} & 
\cellcolor{darkgray!87}\textcolor{white}{0.87} & 
\cellcolor{darkgray!78}\textcolor{white}{0.78} &
\cellcolor{darkgray!39}\textcolor{black}{0.39} \\

Average & 
\cellcolor{darkgray!70}\textcolor{white}{0.70} & 
\cellcolor{darkgray!68}\textcolor{white}{0.68} & 
\cellcolor{darkgray!80}\textcolor{white}{0.80} & 
\cellcolor{darkgray!70}\textcolor{white}{0.70} & 
\cellcolor{darkgray!71}\textcolor{white}{0.71} & 
\cellcolor{darkgray!79}\textcolor{white}{0.79} & 
\cellcolor{darkgray!66}\textcolor{white}{0.66} & 
\cellcolor{darkgray!68}\textcolor{white}{0.68} & 
\cellcolor{darkgray!79}\textcolor{white}{0.79} & 
\cellcolor{darkgray!76}\textcolor{white}{0.76} & 
\cellcolor{darkgray!87}\textcolor{white}{0.67} & 
\cellcolor{darkgray!80}\textcolor{white}{0.80} & 
\cellcolor{darkgray!69}\textcolor{white}{0.69} & 
\cellcolor{darkgray!65}\textcolor{white}{0.65} & 
\cellcolor{darkgray!80}\textcolor{white}{0.80} &
\cellcolor{darkgray!69}\textcolor{white}{0.69} & 
\cellcolor{darkgray!70}\textcolor{white}{0.70} & 
\cellcolor{darkgray!79}\textcolor{white}{0.79} &
\cellcolor{darkgray!36}\textcolor{black}{0.36} \\
  \bottomrule
\end{tabular}
}
\end{table}

\vspace{1mm}
\noindent
\textbf{Software Development.}
The results of our study on automated software development tasks are summarized in \autoref{tab:sd-effectiveness}, which presents a comparative evaluation of seven distinct agent frameworks. The evaluation was conducted using the SRDD dataset, a comprehensive collection of 1,200 instances distributed across five categories: Education (210), Work (240), Life (330), Game (270), and Creation (150).

Our methodology for this assessment was as follows: (1) We benchmarked the seven agent frameworks against the 1,200 simulated software development problems from the SRDD dataset. (2) The generated software projects—comprising Python files, metadata text files, and requirements text files—were quantitatively evaluated for Completeness, Executability, and Consistency using the official evaluation script provided by \chatdev \cite{chatdev}. (3) A category-by-category analysis was performed to identify the specific software development tasks in which each agent excels.

As illustrated in \autoref{tab:sd-effectiveness}, the quality of the software produced by the agents exhibits significant disparities across the three metrics. A key finding is that no single agent framework is omnipotent; each demonstrates a unique profile of strengths and weaknesses. This provides empirical evidence of the current limitations of agent frameworks in code generation, thereby informing future work aimed at their improvement.

A detailed analysis of the metrics reveals the following. For completeness, \agentorchestra demonstrated the highest effectiveness, achieving an impressive average score of 0.86 across the five categories, significantly outperforming the collective average of 0.69. In contrast, \trae registered the poorest effectiveness in completeness, with an average of 0.53. In terms of executability, the \openhands framework proved exceptionally proficient, attaining perfect scores (1.00) in the Game, Education, Life, and Creation categories, and a near-perfect score of 1.00 in the Work category. This outstanding result stands in stark contrast to the overall average executability of 0.79 across all agents.For consistency, \gptswarm emerged as the leader, with a superior average score of 0.85, surpassing the overall agent average of 0.79.Across these five categories, \agentorchestra demonstrated superior performance in completeness, while \openhands excelled in executability, and \gptswarm maintained a leading position in consistency.

To assess overall effectiveness, we utilized the quality score metric proposed by \chatdev, calculated as $Quality = Completeness \times Executability \times Consistency$. Based on this composite metric, \openhands is the top-performing framework for software development tasks, achieving a quality score of 0.47.

Despite the strong effectiveness of \openhands, our findings suggest that its capabilities are not yet sufficient to serve as a universal replacement for all other agents. Each framework possesses distinct advantages, indicating that the landscape of agent-driven software development benefits from a diversity of specialized tools rather than a single, all-encompassing solution.

\begin{table}[t]
\small
  \caption{Comparison of vulnerability detection effectiveness across different vulnerability types on the LLM-SmartAudit benchmark.}
  \label{tab:vd-effectiveness}
  \setlength{\tabcolsep}{1pt}
  \resizebox{\textwidth}{!}{
  \begin{tabular}{lcccccccccc}
    \toprule
    \makecell[c]{\textbf{Subset}\\\textbf{Dataset}} 
    & \makecell[c]{\textbf{Vulnerability Type}\\ \textbf{(\#Contracts)}} 
    & \textbf{\agentorchestra} 
    & \textbf{\owl} 
    & \makecell[c]{\textbf{\seagent}\\ \textbf{(Iter-1)}} 
    & \makecell[c]{\textbf{\seagent}\\ \textbf{(Iter-2)}} 
    & \makecell[c]{\textbf{\seagent}\\ \textbf{(Iter-3)}} 
    & \textbf{\trae} 
    & \textbf{\gptswarm} 
    & \textbf{\openhands} 
    & \textbf{\sweagent} \\
    \midrule
    \multirow{10}{*}{\makecell[c]{Common-\\Vulnerability}}
    & IO (10) 
    & \cellcolor{darkgray!0}\textcolor{black}{0 (0\%)} 
    & \cellcolor{darkgray!10}\textcolor{white}{1 (10\%)} 
    & \cellcolor{darkgray!70}\textcolor{white}{7 (70\%)} 
    & \cellcolor{darkgray!60}\textcolor{white}{6 (60\%)} 
    & \cellcolor{darkgray!90}\textcolor{white}{9 (90\%)} 
    & \cellcolor{darkgray!20}\textcolor{black}{2 (20\%)} 
    & \cellcolor{darkgray!90}\textcolor{white}{9 (90\%)} 
    & \cellcolor{darkgray!80}\textcolor{white}{8 (80\%)} 
    & \cellcolor{darkgray!70}\textcolor{white}{7 (70\%)} \\
    
    & RP (10) 
    & \cellcolor{darkgray!100}\textcolor{white}{10 (100\%)} 
    & \cellcolor{darkgray!100}\textcolor{white}{10 (100\%)} 
    & \cellcolor{darkgray!100}\textcolor{white}{10 (100\%)} 
    & \cellcolor{darkgray!100}\textcolor{white}{10 (100\%)} 
    & \cellcolor{darkgray!100}\textcolor{white}{10 (100\%)} 
    & \cellcolor{darkgray!100}\textcolor{white}{10 (100\%)} 
    & \cellcolor{darkgray!100}\textcolor{white}{10 (100\%)} 
    & \cellcolor{darkgray!100}\textcolor{white}{10 (100\%)} 
    & \cellcolor{darkgray!90}\textcolor{white}{9 (90\%)} \\
    
    & GL (11) 
    & \cellcolor{darkgray!0}\textcolor{black}{0 (0\%)} 
    & \cellcolor{darkgray!0}\textcolor{black}{0 (0\%)} 
    & \cellcolor{darkgray!0}\textcolor{black}{0 (0\%)} 
    & \cellcolor{darkgray!0}\textcolor{black}{0 (0\%)} 
    & \cellcolor{darkgray!9}\textcolor{black}{1 (9\%)} 
    & \cellcolor{darkgray!9}\textcolor{black}{1 (9\%)} 
    & \cellcolor{darkgray!18}\textcolor{black}{2 (18\%)} 
    & \cellcolor{darkgray!9}\textcolor{black}{1 (9\%)} 
    & \cellcolor{darkgray!0}\textcolor{black}{0 (0\%)} \\
    
    & RE (10) 
    & \cellcolor{darkgray!70}\textcolor{white}{7 (70\%)} 
    & \cellcolor{darkgray!100}\textcolor{white}{10 (100\%)} 
    & \cellcolor{darkgray!100}\textcolor{white}{10 (100\%)} 
    & \cellcolor{darkgray!100}\textcolor{white}{10 (100\%)} 
    & \cellcolor{darkgray!100}\textcolor{white}{10 (100\%)} 
    & \cellcolor{darkgray!100}\textcolor{white}{10 (100\%)} 
    & \cellcolor{darkgray!100}\textcolor{white}{10 (100\%)} 
    & \cellcolor{darkgray!100}\textcolor{white}{10 (100\%)} 
    & \cellcolor{darkgray!100}\textcolor{white}{10 (100\%)} \\
    
    & TM (10) 
    & \cellcolor{darkgray!10}\textcolor{black}{1 (10\%)} 
    & \cellcolor{darkgray!30}\textcolor{black}{3 (30\%)} 
    & \cellcolor{darkgray!40}\textcolor{black}{4 (40\%)} 
    & \cellcolor{darkgray!50}\textcolor{white}{5 (50\%)} 
    & \cellcolor{darkgray!40}\textcolor{black}{4 (40\%)} 
    & \cellcolor{darkgray!40}\textcolor{black}{4 (40\%)} 
    & \cellcolor{darkgray!60}\textcolor{white}{6 (60\%)} 
    & \cellcolor{darkgray!10}\textcolor{black}{1 (10\%)} 
    & \cellcolor{darkgray!60}\textcolor{white}{6 (60\%)} \\
    
    & TOD (10) 
    & \cellcolor{darkgray!30}\textcolor{black}{3 (30\%)} 
    & \cellcolor{darkgray!90}\textcolor{white}{9 (90\%)} 
    & \cellcolor{darkgray!90}\textcolor{white}{9 (90\%)} 
    & \cellcolor{darkgray!70}\textcolor{white}{7 (70\%)} 
    & \cellcolor{darkgray!80}\textcolor{white}{8 (80\%)} 
    & \cellcolor{darkgray!90}\textcolor{white}{9 (90\%)} 
    & \cellcolor{darkgray!80}\textcolor{white}{8 (80\%)} 
    & \cellcolor{darkgray!40}\textcolor{black}{4 (40\%)} 
    & \cellcolor{darkgray!80}\textcolor{white}{8 (80\%)} \\
    
    & USE (10) 
    & \cellcolor{darkgray!0}\textcolor{black}{0 (0\%)} 
    & \cellcolor{darkgray!20}\textcolor{black}{2 (20\%)} 
    & \cellcolor{darkgray!80}\textcolor{white}{8 (80\%)} 
    & \cellcolor{darkgray!80}\textcolor{white}{8 (80\%)} 
    & \cellcolor{darkgray!80}\textcolor{white}{8 (80\%)} 
    & \cellcolor{darkgray!20}\textcolor{black}{2 (20\%)} 
    & \cellcolor{darkgray!90}\textcolor{white}{9 (90\%)} 
    & \cellcolor{darkgray!30}\textcolor{black}{3 (30\%)} 
    & \cellcolor{darkgray!90}\textcolor{white}{9 (90\%)} \\
    
    & TX (10) 
    & \cellcolor{darkgray!70}\textcolor{white}{7 (70\%)} 
    & \cellcolor{darkgray!70}\textcolor{white}{7 (70\%)} 
    & \cellcolor{darkgray!70}\textcolor{white}{7 (70\%)} 
    & \cellcolor{darkgray!70}\textcolor{white}{7 (70\%)} 
    & \cellcolor{darkgray!70}\textcolor{white}{7 (70\%)} 
    & \cellcolor{darkgray!60}\textcolor{white}{6 (60\%)} 
    & \cellcolor{darkgray!70}\textcolor{white}{7 (70\%)} 
    & \cellcolor{darkgray!60}\textcolor{white}{6 (60\%)} 
    & \cellcolor{darkgray!80}\textcolor{white}{8 (80\%)} \\
    
    & UD (10) 
    & \cellcolor{darkgray!90}\textcolor{white}{9 (90\%)} 
    & \cellcolor{darkgray!80}\textcolor{white}{8 (80\%)} 
    & \cellcolor{darkgray!100}\textcolor{white}{10 (100\%)} 
    & \cellcolor{darkgray!100}\textcolor{white}{10 (100\%)} 
    & \cellcolor{darkgray!100}\textcolor{white}{10 (100\%)} 
    & \cellcolor{darkgray!100}\textcolor{white}{10 (100\%)} 
    & \cellcolor{darkgray!100}\textcolor{white}{10 (100\%)} 
    & \cellcolor{darkgray!100}\textcolor{white}{10 (100\%)} 
    & \cellcolor{darkgray!90}\textcolor{white}{9 (90\%)} \\
    
    & USU (11) 
    & \cellcolor{darkgray!100}\textcolor{white}{11 (100\%)} 
    & \cellcolor{darkgray!91}\textcolor{white}{10 (91\%)} 
    & \cellcolor{darkgray!100}\textcolor{white}{11 (100\%)} 
    & \cellcolor{darkgray!91}\textcolor{white}{10 (91\%)} 
    & \cellcolor{darkgray!91}\textcolor{white}{10 (91\%)} 
    & \cellcolor{darkgray!91}\textcolor{white}{10 (91\%)} 
    & \cellcolor{darkgray!91}\textcolor{white}{10 (91\%)} 
    & \cellcolor{darkgray!55}\textcolor{white}{6 (55\%)} 
    & \cellcolor{darkgray!100}\textcolor{white}{11 (100\%)} \\

    \midrule
    \multirow{5}{*}{CVE}  
    & Access Control (4) 
    & \cellcolor{darkgray!0}\textcolor{black}{0 (0\%)} 
    & \cellcolor{darkgray!100}\textcolor{white}{1 (25\%)} 
    & \cellcolor{darkgray!50}\textcolor{white}{2 (50\%)} 
    & \cellcolor{darkgray!75}\textcolor{white}{3 (75\%)} 
    & \cellcolor{darkgray!50}\textcolor{white}{2 (50\%)} 
    & \cellcolor{darkgray!50}\textcolor{white}{2 (50\%)} 
    & \cellcolor{darkgray!100}\textcolor{white}{4 (100\%)} 
    & \cellcolor{darkgray!100}\textcolor{white}{4 (100\%)} 
    & \cellcolor{darkgray!100}\textcolor{white}{4 (100\%)} \\
    
    & Overflow (3) 
    & \cellcolor{darkgray!33}\textcolor{black}{1 (33\%)} 
    & \cellcolor{darkgray!100}\textcolor{white}{3 (100\%)} 
    & \cellcolor{darkgray!100}\textcolor{white}{3 (100\%)} 
    & \cellcolor{darkgray!100}\textcolor{white}{3 (100\%)} 
    & \cellcolor{darkgray!100}\textcolor{white}{3 (100\%)} 
    & \cellcolor{darkgray!100}\textcolor{white}{3 (100\%)} 
    & \cellcolor{darkgray!100}\textcolor{white}{3 (100\%)} 
    & \cellcolor{darkgray!100}\textcolor{white}{3 (100\%)} 
    & \cellcolor{darkgray!100}\textcolor{white}{3 (100\%)} \\
    
    & Logic Error (4) 
    & \cellcolor{darkgray!0}\textcolor{black}{0 (0\%)} 
    & \cellcolor{darkgray!0}\textcolor{black}{0 (0\%)} 
    & \cellcolor{darkgray!0}\textcolor{black}{0 (0\%)} 
    & \cellcolor{darkgray!0}\textcolor{black}{0 (0\%)} 
    & \cellcolor{darkgray!0}\textcolor{black}{0 (0\%)} 
    & \cellcolor{darkgray!0}\textcolor{black}{0 (0\%)} 
    & \cellcolor{darkgray!0}\textcolor{black}{0 (0\%)} 
    & \cellcolor{darkgray!25}\textcolor{black}{1 (25\%)} 
    & \cellcolor{darkgray!25}\textcolor{black}{1 (25\%)} \\
    
    & Delegatecall (1) 
    & \cellcolor{darkgray!100}\textcolor{white}{1 (100\%)} 
    & \cellcolor{darkgray!100}\textcolor{white}{1 (100\%)} 
    & \cellcolor{darkgray!100}\textcolor{white}{1 (100\%)} 
    & \cellcolor{darkgray!100}\textcolor{white}{1 (100\%)} 
    & \cellcolor{darkgray!100}\textcolor{white}{1 (100\%)} 
    & \cellcolor{darkgray!100}\textcolor{white}{1 (100\%)} 
    & \cellcolor{darkgray!100}\textcolor{white}{1 (100\%)} 
    & \cellcolor{darkgray!100}\textcolor{white}{1 (100\%)} 
    & \cellcolor{darkgray!100}\textcolor{white}{1 (100\%)} \\
    
    & Bad Randomness (1) 
    & \cellcolor{darkgray!100}\textcolor{white}{1 (100\%)} 
    & \cellcolor{darkgray!100}\textcolor{white}{1 (100\%)} 
    & \cellcolor{darkgray!100}\textcolor{white}{1 (100\%)} 
    & \cellcolor{darkgray!100}\textcolor{white}{1 (100\%)} 
    & \cellcolor{darkgray!100}\textcolor{white}{1 (100\%)} 
    & \cellcolor{darkgray!100}\textcolor{white}{1 (100\%)} 
    & \cellcolor{darkgray!100}\textcolor{white}{1 (100\%)} 
    & \cellcolor{darkgray!0}\textcolor{black}{0 (0\%)} 
    & \cellcolor{darkgray!100}\textcolor{white}{1 (100\%)} \\

    \midrule
    & \textbf{Total (115)}
    & \cellcolor{darkgray!44}\textcolor{black}{\textbf{51} (44\%)} 
    & \cellcolor{darkgray!59}\textcolor{white}{\textbf{66} (57\%)} 
    & \cellcolor{darkgray!72}\textcolor{white}{\textbf{83} (72\%)} 
    & \cellcolor{darkgray!70}\textcolor{white}{\textbf{81} (70\%)} 
    & \cellcolor{darkgray!73}\textcolor{white}{\textbf{84} (73\%)} 
    & \cellcolor{darkgray!62}\textcolor{white}{\textbf{71} (62\%)} 
    & \cellcolor{darkgray!78}\textcolor{white}{\textbf{80} (78\%)} 
    & \cellcolor{darkgray!59}\textcolor{white}{\textbf{68} (59\%)} 
    & \cellcolor{darkgray!76}\textcolor{white}{\textbf{87} (76\%)} \\
  \bottomrule
\end{tabular}
}
\end{table}

\vspace{1mm}
\noindent
\textbf{Vulnerability Detection.} \autoref{tab:vd-effectiveness} shows the experimental results of seven agent frameworks on vulnerability detection task.
The common-vulnerability subset dataset contains 10 types of vulnerabilities, specifically Reentrancy (RE), Integer Overflow/Underflow (IO), Unchecked send (USE), Unsafe Delegatecall (UD), Transaction Order Dependence (TOD), Time Manipulation (TM), Randomness Prediction (RP), Authorization Issue using ‘tx.origin’ (TX), Unsafe Suicide (USU), and Gas Limitation (GL). The CVE subset dataset contains 5 types of vulnerabilities, including Access Control, Overflow, Logic Error, Delegatecall, and Bad Randomness.

We adopt the following methods to answer this question: (1) We evaluate seven agent frameworks in our assessment on the 115 vulnerable smart contracts across different vulnerability types. (2) When the vulnerability detection report generated by the agent framework is compared with the ground truth of the labeled dataset, the detection accuracy rate for each vulnerability type can be calculated. (3) Through type-by-type analysis, we can observe which vulnerability detection tasks these agents perform well in. 


For the research results of vulnerability detection presented in \autoref{tab:vd-effectiveness}, which shows the vulnerability detection effectiveness of seven agent frameworks across 5 types of CVE dataset and 10 types of labeled dataset. When the agent framework generates a vulnerability detection report, it details whether vulnerabilities exist and the types of vulnerabilities present. Each row in \autoref{tab:vd-effectiveness} represents a vulnerability detection type, and each cell represents the number of vulnerabilities detected by the agent framework for this vulnerability type, along with the percentage relative to the total number of vulnerabilities.

\autoref{tab:vd-effectiveness} shows that the accuracy rates of smart contract vulnerability detection across various agents are slightly similar. In fact, for the GL vulnerability type, all seven agents demonstrated very low detection efficiency, basically failing to discover any vulnerabilities. However, for vulnerability types such as PR and UD, the results are very significant, approaching a detection accuracy rate of 100\%.
We observe that the seven agents demonstrate comparable performance in the vulnerability detection task, with \gptswarm achieving the highest score of 77\%, \agentorchestra recording the lowest at 44\%, and \seagent in Iter-2 representing the median at 70\%. However, they have not broken through 90\%, which indicates that the seven agent frameworks we evaluated still have room for improvement.


\begin{table}[t]
\small
  \caption{Comparison of program repair effectiveness across different repositories on the SWE-bench Lite benchmark.}
  \label{tab:pr-effectiveness}
  \setlength{\tabcolsep}{1pt}
  \resizebox{\textwidth}{!}{
  \begin{tabular}{lcccccccccc}
    \toprule
    \makecell[c]{\textbf{Repository}\\\textbf{(\#Issues)}} & 
    \textbf{\agentorchestra} & \textbf{\owl} & 
    \makecell[c]{\textbf{\seagent}\\ \textbf{(Iter-1)}} & 
    \makecell[c]{\textbf{\seagent}\\ \textbf{(Iter-2)}} & 
    \makecell[c]{\textbf{\seagent}\\ \textbf{(Iter-3)}} & 
    \textbf{\trae} & \textbf{\gptswarm} & 
    \textbf{\openhands} & \textbf{\sweagent} \\
    \midrule
    \codeff{astropy} (6) &
    \cellcolor{darkgray!0}\textcolor{black}{0 (0\%)} & 
    \cellcolor{darkgray!17}\textcolor{black}{1 (17\%)} & 
    \cellcolor{darkgray!33}\textcolor{black}{2 (33\%)} & 
    \cellcolor{darkgray!50}\textcolor{white}{3 (50\%)} & 
    \cellcolor{darkgray!33}\textcolor{black}{2 (33\%)} & 
    \cellcolor{darkgray!50}\textcolor{white}{3 (50\%)} & 
    \cellcolor{darkgray!0}\textcolor{black}{0 (0\%)} & 
    \cellcolor{darkgray!50}\textcolor{white}{3 (50\%)} & 
    \cellcolor{darkgray!33}\textcolor{black}{2 (33\%)} 
    \\
    
    \codeff{django} (114) &
    \cellcolor{darkgray!6}\textcolor{black}{7 (6\%)} & 
    \cellcolor{darkgray!6}\textcolor{black}{7 (6\%)} & 
    \cellcolor{darkgray!52}\textcolor{white}{59 (52\%)} & 
    \cellcolor{darkgray!59}\textcolor{white}{67 (59\%)} & 
    \cellcolor{darkgray!60}\textcolor{white}{68 (60\%)} & 
    \cellcolor{darkgray!60}\textcolor{white}{68 (60\%)} & 
    \cellcolor{darkgray!12}\textcolor{black}{14 (12\%)} & 
    \cellcolor{darkgray!54}\textcolor{white}{61 (54\%)} & 
    \cellcolor{darkgray!58}\textcolor{white}{66 (58\%)} 
    \\
    
    \codeff{flask} (3) &
    \cellcolor{darkgray!0}\textcolor{black}{0 (0\%)} & 
    \cellcolor{darkgray!0}\textcolor{black}{0 (0\%)} & 
    \cellcolor{darkgray!0}\textcolor{black}{0 (0\%)} & 
    \cellcolor{darkgray!0}\textcolor{black}{0 (0\%)} & 
    \cellcolor{darkgray!0}\textcolor{black}{0 (0\%)} & 
    \cellcolor{darkgray!0}\textcolor{black}{0 (0\%)} & 
    \cellcolor{darkgray!0}\textcolor{black}{0 (0\%)} & 
    \cellcolor{darkgray!0}\textcolor{black}{0 (0\%)} & 
    \cellcolor{darkgray!0}\textcolor{black}{0 (0\%)} 
    \\
    
    \codeff{matplotlib} (23) &
    \cellcolor{darkgray!1}\textcolor{black}{0 (0\%)} & 
    \cellcolor{darkgray!9}\textcolor{black}{2 (9\%)} & 
    \cellcolor{darkgray!39}\textcolor{black}{9 (39\%)} & 
    \cellcolor{darkgray!61}\textcolor{white}{14 (61\%)} & 
    \cellcolor{darkgray!61}\textcolor{white}{14 (61\%)} & 
    \cellcolor{darkgray!48}\textcolor{black}{11 (48\%)} & 
    \cellcolor{darkgray!0}\textcolor{black}{0 (0\%)} & 
    \cellcolor{darkgray!48}\textcolor{black}{11 (48\%)} & 
    \cellcolor{darkgray!61}\textcolor{white}{14 (61\%)} 
    \\
    
    \codeff{seaborn} (4) &
    \cellcolor{darkgray!0}\textcolor{black}{0 (0\%)} & 
    \cellcolor{darkgray!25}\textcolor{black}{1 (25\%)} & 
    \cellcolor{darkgray!75}\textcolor{white}{3 (75\%)} & 
    \cellcolor{darkgray!100}\textcolor{white}{4 (100\%)} & 
    \cellcolor{darkgray!75}\textcolor{white}{3 (75\%)} & 
    \cellcolor{darkgray!75}\textcolor{white}{3 (75\%)} & 
    \cellcolor{darkgray!0}\textcolor{black}{0 (0\%)} & 
    \cellcolor{darkgray!75}\textcolor{white}{3 (75\%)} & 
    \cellcolor{darkgray!100}\textcolor{white}{4 (100\%)} 
    \\
    
    \codeff{pylint} (6) &
    \cellcolor{darkgray!0}\textcolor{black}{0 (0\%)} & 
    \cellcolor{darkgray!0}\textcolor{black}{0 (0\%)} & 
    \cellcolor{darkgray!33}\textcolor{black}{2 (33\%)} & 
    \cellcolor{darkgray!50}\textcolor{white}{3 (50\%)} & 
    \cellcolor{darkgray!33}\textcolor{black}{2 (33\%)} & 
    \cellcolor{darkgray!50}\textcolor{white}{3 (50\%)} & 
    \cellcolor{darkgray!0}\textcolor{black}{0 (0\%)} & 
    \cellcolor{darkgray!50}\textcolor{black}{2 (33\%)} & 
    \cellcolor{darkgray!50}\textcolor{white}{3 (50\%)} 
    \\
    
    \codeff{pytest} (17) &
    \cellcolor{darkgray!0}\textcolor{black}{0 (0\%)} & 
    \cellcolor{darkgray!6}\textcolor{black}{1 (6\%)} & 
    \cellcolor{darkgray!35}\textcolor{black}{6 (35\%)} & 
    \cellcolor{darkgray!35}\textcolor{black}{6 (35\%)} & 
    \cellcolor{darkgray!35}\textcolor{black}{6 (35\%)} & 
    \cellcolor{darkgray!35}\textcolor{black}{6 (35\%)} & 
    \cellcolor{darkgray!0}\textcolor{black}{0 (0\%)} & 
    \cellcolor{darkgray!47}\textcolor{black}{8 (47\%)} & 
    \cellcolor{darkgray!29}\textcolor{black}{5 (29\%)} 
    \\
    
    \codeff{requests} (6) &
    \cellcolor{darkgray!17}\textcolor{black}{1 (17\%)} & 
    \cellcolor{darkgray!33}\textcolor{black}{2 (33\%)} & 
    \cellcolor{darkgray!17}\textcolor{black}{1 (17\%)} & 
    \cellcolor{darkgray!17}\textcolor{black}{1 (17\%)} & 
    \cellcolor{darkgray!50}\textcolor{white}{3 (50\%)} & 
    \cellcolor{darkgray!50}\textcolor{white}{3 (50\%)} & 
    \cellcolor{darkgray!0}\textcolor{black}{0 (0\%)} & 
    \cellcolor{darkgray!0}\textcolor{black}{0 (0\%)} & 
    \cellcolor{darkgray!67}\textcolor{white}{4 (67\%)} 
    \\
    
    \codeff{scikit} (23) &
    \cellcolor{darkgray!0}\textcolor{black}{0 (0\%)} & 
    \cellcolor{darkgray!4}\textcolor{black}{1 (4\%)} & 
    \cellcolor{darkgray!74}\textcolor{white}{17 (74\%)} & 
    \cellcolor{darkgray!70}\textcolor{white}{16 (70\%)} & 
    \cellcolor{darkgray!65}\textcolor{white}{15 (65\%)} & 
    \cellcolor{darkgray!61}\textcolor{white}{14 (61\%)} & 
    \cellcolor{darkgray!4}\textcolor{black}{1 (4\%)} & 
    \cellcolor{darkgray!70}\textcolor{white}{16 (70\%)} & 
    \cellcolor{darkgray!70}\textcolor{white}{16 (70\%)} 
    \\
    
    \codeff{sphinx} (16) &
    \cellcolor{darkgray!6}\textcolor{black}{1 (6\%)} & 
    \cellcolor{darkgray!6}\textcolor{black}{1 (6\%)} & 
    \cellcolor{darkgray!44}\textcolor{black}{7 (44\%)} & 
    \cellcolor{darkgray!56}\textcolor{white}{9 (56\%)} & 
    \cellcolor{darkgray!56}\textcolor{white}{9 (56\%)} & 
    \cellcolor{darkgray!50}\textcolor{white}{8 (50\%)} & 
    \cellcolor{darkgray!0}\textcolor{black}{0 (0\%)} & 
    \cellcolor{darkgray!19}\textcolor{black}{3 (19\%)} & 
    \cellcolor{darkgray!56}\textcolor{white}{9 (56\%)} 
    \\
    
    \codeff{sympy} (77) &
    \cellcolor{darkgray!0}\textcolor{black}{0 (0\%)} & 
    \cellcolor{darkgray!19}\textcolor{black}{15 (19\%)} & 
    \cellcolor{darkgray!44}\textcolor{black}{34 (44\%)} & 
    \cellcolor{darkgray!44}\textcolor{black}{34 (44\%)} & 
    \cellcolor{darkgray!48}\textcolor{black}{37 (48\%)} & 
    \cellcolor{darkgray!43}\textcolor{black}{33 (43\%)} & 
    \cellcolor{darkgray!0}\textcolor{black}{0 (0\%)} & 
    \cellcolor{darkgray!48}\textcolor{black}{37 (48\%)} & 
    \cellcolor{darkgray!44}\textcolor{black}{34 (44\%)} 
    \\
    
    \codeff{xarray} (5) &
    \cellcolor{darkgray!0}\textcolor{black}{0 (0\%)} & 
    \cellcolor{darkgray!0}\textcolor{black}{0 (0\%)} & 
    \cellcolor{darkgray!40}\textcolor{black}{2 (40\%)} & 
    \cellcolor{darkgray!40}\textcolor{black}{2 (40\%)} & 
    \cellcolor{darkgray!40}\textcolor{black}{2 (40\%)} & 
    \cellcolor{darkgray!40}\textcolor{black}{2 (40\%)} & 
    \cellcolor{darkgray!0}\textcolor{black}{0 (0\%)} & 
    \cellcolor{darkgray!40}\textcolor{black}{2 (40\%)} & 
    \cellcolor{darkgray!40}\textcolor{black}{2 (40\%)} 
    \\
    \midrule
    \textbf{Total (300)} & 
    \cellcolor{darkgray!3}\textcolor{black}{\textbf{10 (3\%)}} & 
    \cellcolor{darkgray!10}\textcolor{black}{\textbf{31 (10\%)}} & 
    \cellcolor{darkgray!47}\textcolor{black}{\textbf{142 (47\%)}} & 
    \cellcolor{darkgray!53}\textcolor{white}{\textbf{159 (53\%)}} & 
    \cellcolor{darkgray!54}\textcolor{white}{\textbf{161 (54\%)}} & 
    \cellcolor{darkgray!51}\textcolor{white}{\textbf{154 (51\%)}} & 
    \cellcolor{darkgray!5}\textcolor{black}{\textbf{15 (5\%)}} & 
    \cellcolor{darkgray!49}\textcolor{black}{\textbf{146 (49\%)}} & 
    \cellcolor{darkgray!53}\textcolor{white}{\textbf{159 (53\%)}} 
    \\
    \bottomrule
\end{tabular}
}
\end{table}

\vspace{1mm}
\noindent
\textbf{Program Repair.} The SWEBench-Lite benchmark comprises 12 GitHub repositories with a total of 300 instances. Our methodology for the program repair task was as follows: (1) we evaluated the seven agent frameworks on the 300 real-world issues spanning the 12 repositories; (2) the agents generated patches in diff format, which we standardized into the official JSON schema; (3) we further stratified the 300 instances by repository type across the 12 GitHub projects—for example, \codeff{django} is a web development framework repository with 114 issues; and (4) by analyzing results repository-by-repository, we identified which issues were successfully repaired by each agent. 

The detailed results for program repair are summarized in \autoref{tab:pr-effectiveness}, which reports the number of successfully repaired issues per agent framework for each of the 12 repositories. A repair is considered successful if the diff-format patch generated by an agent passes the official test suite. In \autoref{tab:pr-effectiveness}, each row corresponds to a repository, and each cell represents the number of issues successfully repaired by a specific agent within that repository.

\autoref{tab:pr-effectiveness} reveals substantial differences in accuracy across agents. Notably, \seagent(Iter-3) achieved the highest accuracy, repairing 54\% of the total issues., 
whereas the average success rate across all agents was only 36\%.
Additionally, for the \codeff{django} repository, which contains the largest number of issues (114), \trae and \seagent (Iter-3) achieved the best performance, each successfully repairing 68 issues. 
In contrast, for \codeff{flask} repository, which contains the fewest issues (3), none of the agents succeeded in repairing any issue, an unexpected outcome suggesting that the number of issues does not necessarily correlate with task difficulty.
Note that \agentorchestra, \owl, and \gptswarm exhibit distinctly poor effectiveness. The underlying reason is that they do not leverage Patch tooling \cite{sweagent}, which leads to failures in generating correct diff-format patches. This limitation reflects a broader challenge inherent to current LLM capabilities.

In total, \seagent(Iter-2) and \sweagent fully repaired (100
This indicates that agent frameworks are not universally effective for program repair. Although overall effectiveness was modest, these findings provide concrete evidence of the limitations of current agent frameworks in program repair and offer actionable insights for future improvements.

\begin{tcolorbox}[size=title]
\textbf{Answer to RQ1.} The key takeaways are as follows:
\begin{itemize}
    \item \textbf{Software Development:} A clear effectiveness trade-off is observed among completeness, executability, and consistency. \openhands achieves the best overall balance, with the highest quality score of 0.47.
    \item \textbf{Vulnerability Detection:} The average success rate across the seven agent frameworks is approximately 66\%, with \gptswarm achieving the highest detection accuracy of 77\%. Nevertheless, this performance remains suboptimal and leaves considerable room for improvement.
    \item \textbf{Program Repair:} Effectiveness is modest, with at most roughly half of the issues repaired by only 4 out of 7 agents, while the rest achieve low repair rates, highlighting significant potential for improvement.
\end{itemize}
\end{tcolorbox}
\subsection{Efficiency of Agent Frameworks (RQ2)}
\label{subsec:rq2}

To address RQ2, we analyze the efficiency of the seven agent frameworks in the same set of code-centric software engineering tasks introduced in \S~\ref{subsec:rq1}. Unlike RQ1, which focuses solely on final outcomes, RQ2 delves into the agents' execution processes, providing a more nuanced understanding of their operational behavior. 
In contrast to traditional software engineering automation tools, which execute pre-defined, statically encoded procedures, agents operate in a dynamic and adaptive manner. Consequently, assessing the efficiency of these processes is essential for gaining insights into the level of intelligence and adaptability demonstrated by each framework.

\begin{table}[t]
\small
  \caption{Comparison of software development efficiency across different agent frameworks.}
  \vspace{-3mm}
  \label{tab:sd-efficiency}
  \setlength{\tabcolsep}{1pt}
  \resizebox{\textwidth}{!}{
  \begin{threeparttable}
  \begin{tabular}{lcccccccccc}
    \toprule
    \textbf{Metric}
    & \textbf{\agentorchestra} 
    & \textbf{\owl} 
    & \makecell[c]{\textbf{\seagent}\\ \textbf{(Iter-1)}} 
    & \makecell[c]{\textbf{\seagent}\\ \textbf{(Iter-2)}} 
    & \makecell[c]{\textbf{\seagent}\\ \textbf{(Iter-3)}} 
    & \textbf{\trae} 
    & \textbf{\gptswarm} 
    & \textbf{\openhands} 
    & \textbf{\sweagent}
    & \textbf{Average}
    \\
    \midrule
    \textbf{Average Trajectory Steps} & 40.20 & 19.90 & 43.90 & 3.70 & 3.20 
    & 28.20 & 7.00 & \textbf{81.28} & 27.40 & 28.31\\
    \textbf{Correction Attempts} & \textbf{16.7} & 4.20 & 3.60 & 1.10 & 0.70 
    & 6.10 & 1.21 & 7.79 & 1.90 & 4.81 \\  
    \textbf{Correction Rate} & \textbf{41.54\%} & 21.10\% & 8.20\% & 29.73\% & 21.86\% 
    & 21.63\% & 17.19\% & 9.58\% & 6.93\% & 16.99\% \\
  \bottomrule
\end{tabular}
     
    \end{threeparttable}
}

\end{table}

\vspace{1mm}
\noindent
\textbf{Software Development.}
\autoref{tab:sd-efficiency} presents the execution trajectory analysis for the software development task, revealing distinct operational characteristics across different agent frameworks. Statistical analysis shows that seven frameworks exhibit an average trajectory length of 28.31 steps, with \openhands recording the longest trajectory of 81.28 steps. 
Note that \seagent exhibits a distinctive iterative pattern: the first iteration comprises 43.9 steps, while the second and third iterations show markedly shorter trajectories of 3.70 and 3.20 steps, respectively. This progressive reduction suggests that \seagent employs an adaptive refinement strategy, in which an initial comprehensive exploration is followed by focused, targeted adjustments in subsequent iterations. Such a pattern reflects efficient convergence behavior, as the agent leverages insights from earlier iterations to streamline problem-solving in later stages.

Our analysis reveals that \agentorchestra exhibits the highest correction attempts at 16.7, accounting for approximately 41.54\% of its trajectory length. This substantially exceeds the average correction attempts of 4.81 and correction rate of 16.99\%, representing the highest values among all evaluated agents. 
In contrast, \sweagent achieves the lowest correction rate among all agents at merely 6.93\%, while \gptswarm requires the fewest correction attempts at 1.21.
Notably, \seagent (Iter-1) initially exhibits the lowest correction rate at 8.20\%. However, an intriguing pattern emerges across subsequent iterations: although the absolute number of correction attempts decreases to 1.10 in the second iteration and 0.70 in the third iteration, the correction rate paradoxically increases. This phenomenon is attributable to the substantial reduction in trajectory length across iterations, resulting in a higher proportional correction rate despite fewer absolute corrections. 


\begin{table}[t]
\small
  \caption{Comparison of vulnerability detection efficiency across different agent frameworks.}
  \vspace{-3mm}
  \label{tab:vd-efficiency}
  \setlength{\tabcolsep}{1pt}
  \resizebox{\textwidth}{!}{
  \begin{threeparttable}
  \begin{tabular}{lcccccccccc}
    \toprule
    \textbf{Metric}
    & \textbf{\agentorchestra} 
    & \textbf{\owl} 
    & \makecell[c]{\textbf{\seagent}\\ \textbf{(Iter-1)}} 
    & \makecell[c]{\textbf{\seagent}\\ \textbf{(Iter-2)}} 
    & \makecell[c]{\textbf{\seagent}\\ \textbf{(Iter-3)}} 
    & \textbf{\trae} 
    & \textbf{\gptswarm} 
    & \textbf{\openhands} 
    & \textbf{\sweagent}
    & \textbf{Average}
    \\
    \midrule
    \textbf{Average Trajectory Steps} 
    & \textbf{46.50} & 7.50 & 1.00 & 1.40 & 7.00
    & 8.90 & 1.00 & 11.00 & 9.50 & 10.40\\
    \textbf{Correction Attempts} 
    & \textbf{21.04} & 0.80 & 0.00 & 0.10 & 1.30
    & 2.10 & 0.00 & 3.70 & 0.50 & 3.28\\
    \textbf{Correction Rate} 
    & \textbf{45.25\%} & 10.67\% & 0.00\% & 7.14\% & 18.57\% 
    & 23.60\% & 0.00\% & 33.64\% & 5.26\% & 31.54\%\\
  \bottomrule
\end{tabular}
     
    \end{threeparttable}
}
\end{table}

\vspace{1mm}
\noindent
\textbf{Vulnerability Detection.}
\autoref{tab:vd-efficiency} presents the trajectory statistics of these agents in the vulnerability detection task. The analysis reveals that seven frameworks exhibit an average trajectory length of 10.40 steps, with \agentorchestra demonstrating the longest trajectory at 46.5 steps. In contrast, \seagent demonstrate the shortest trajectories, with the third iteration reaching only 7.00 steps.
Furthermore, \autoref{tab:vd-efficiency} presents the correction attempts and corresponding correction rates for each agent framework in the vulnerability detection task. 
\agentorchestra exhibits 21.04 correction attempts, constituting nearly 45.25\% of its total trajectory steps, which represents the highest correction-to-execution ratio among the evaluated systems. This substantially exceeds the average of 3.28 correction attempts and 31.54\% correction rate. 
Moreover, both \seagent (Iter-1) and \gptswarm exhibit zero correction attempts, representing the most efficient execution patterns among all evaluated agents.


\begin{table}[t]
\small
  \caption{Comparison of program repair efficiency across different agent frameworks.}
  \vspace{-3mm}
  \label{tab:pr-efficiency}
  \setlength{\tabcolsep}{1pt}
  \resizebox{\textwidth}{!}{
  \begin{threeparttable}
  \begin{tabular}{lcccccccccc}
    \toprule
    \textbf{Metric}
    & \textbf{\agentorchestra} 
    & \textbf{\owl} 
    & \makecell[c]{\textbf{\seagent}\\ \textbf{(Iter-1)}} 
    & \makecell[c]{\textbf{\seagent}\\ \textbf{(Iter-2)}} 
    & \makecell[c]{\textbf{\seagent}\\ \textbf{(Iter-3)}} 
    & \textbf{\trae} 
    & \textbf{\gptswarm} 
    & \textbf{\openhands} 
    & \textbf{\sweagent}
    & \textbf{Average}
    \\
    \midrule
    \textbf{Average Trajectory Steps} 
    & 46.80 & 64.50 & 51.50 & 67.40 & 69.10 
    & \textbf{78.10} & 2.90 & 69.30 & 67.40 & 57.44 \\
    \textbf{Correction Attempts} & 21.10 & 1.00 & 6.70 & 9.90 & 9.50
    & 19.50 & 0.50 & \textbf{25.20} & 10.40 & 11.53 \\
    \textbf{Correction Rate} & \textbf{45.09\%} & 1.55\% & 13.01\% & 14.69\% & 13.75\%
    & 24.97\% & 17.24\% & 36.36\% & 15.43\% & 20.08\% \\
  \bottomrule
\end{tabular}
     
    \end{threeparttable}
}
\end{table}

\vspace{1mm}
\noindent
\textbf{Program Repair.} 
\autoref{tab:pr-efficiency} presents the execution trajectory analysis for the program repair task. 
Statistical analysis demonstrates that seven frameworks exhibit an average trajectory length of 57.44 steps, 
\trae records the longest average trajectory of 78.1 steps among single-agent frameworks, warranting a detailed examination of its execution pattern. Our analysis reveals that the repair process of \trae consists of: (1) 41.1 Bash commands for environment configuration, code execution, and repair validation; (2) 30 file editing operations; (3) 6 problem analysis steps; and (4) 1 completion command, totaling 78.1 steps. This decomposition indicates that \trae's extended trajectory primarily stems from its more comprehensive problem analysis phase and increased reliance on environment validation commands, distinguishing it from other frameworks that adopt more streamlined execution strategies.
Moreover, since \gptswarm employs the CodeReact \cite{gptswarm} mechanism with a predetermined constraint of exactly 3 React cycles, it exhibits the shortest trajectory length of 2.90 steps among all seven evaluated agent frameworks, representing a distinct operational paradigm that prioritizes execution efficiency through fixed iteration constraints rather than adaptive exploration.

Although \trae exhibits the longest trajectory among all agents, its correction attempts of 19.50 do not represent the maximum value. \openhands records the highest correction attempts at 25.20, accounting for 36.36\% of its trajectory length, substantially exceeding averages of 11.53 attempts at 20.08\% rate. 
In contrast, \gptswarm demonstrates the fewest correction attempts at merely 0.5, while \owl exhibits the lowest correction rate at only 1.55\%. However, these minimal correction metrics do not translate to superior performance in program repair tasks. Both \gptswarm and \owl fail to identify errors in their self-generated content, suggesting that their low correction frequencies reflect inadequate self-monitoring capabilities rather than execution efficiency. This observation underscores a critical insight: the absence of correction attempts may indicate a deficiency in error detection mechanisms rather than optimal performance. This deficiency can be attributed to the fact that neither \gptswarm nor \owl incorporates version control tools such as \texttt{Git}, utilized by \sweagent, or employs specialized diff format validation utilities like \texttt{unidiff}. The absence of these essential tools significantly undermines their operational efficiency and error detection capabilities.



\begin{tcolorbox}[size=title]
\textbf{Answer to RQ2.} The key takeaways are as follows:
\begin{itemize}
    \item \textbf{Trajectory Steps:} 
    \openhands, \agentorchestra, and \seagent(Iter-3) record the longest trajectories in software development, vulnerability detection, and program repair, respectively.
    \item \textbf{Correction Attempts:} \agentorchestra shows the longest correction attempts in software development and vulnerability detection, while \openhands records the longest in program repair.
    \item \textbf{Correction Rate:} Across the three tasks, \agentorchestra achieves the highest correction rate, reflecting its lower efficiency due to frequent revisions.
\end{itemize}
\end{tcolorbox}
\subsection{Overhead of Agent Frameworks (RQ3)}

\begin{table}[t]
\small
  \caption{Comparison of monetary overhead (USD) across different agent frameworks on three tasks.}
  \label{tab:cost-sd-vd-pr}
  \setlength{\tabcolsep}{1pt}
  \resizebox{\textwidth}{!}{
  \begin{tabular}{lcccccccccc}
    \toprule
    \textbf{Task} 
    & \textbf{\agentorchestra} 
    & \textbf{\owl} 
    & \makecell[c]{\textbf{\seagent}\\ \textbf{(Iter-1)}} 
    & \makecell[c]{\textbf{\seagent}\\ \textbf{(Iter-2)}} 
    & \makecell[c]{\textbf{\seagent}\\ \textbf{(Iter-3)}} 
    & \textbf{\trae} 
    & \textbf{\gptswarm} 
    & \textbf{\openhands} 
    & \textbf{\sweagent}
    & \textbf{Total}\\
    \midrule
    \textbf{Software Development} & \textbf{\$292.01} & \$21.05 & \$43.10 & \$31.01 & \$25.37 & \$48.90 & \$13.49 & \$39.98 & \$29.99 & \textbf{\$544.90}\\
    \textbf{Vulnerability Detection} & \$14.13 & \$1.89 & \$0.29 & \$0.26 & \$1.27 & \$1.91 & \$0.67 & \textbf{\$20.91} & \$1.07 & \$42.40\\
    \textbf{Program Repair} & \textbf{\$64.05} & \$2.49 & \$27.78 & \$38.40 & \$40.73 & \$15.53 & \$2.13 & \$54.48 & \$42.16 & \$287.75\\
    \midrule
    \textbf{Total} & \textbf{\$370.19} & \$25.43 & \multicolumn{3}{c}{\$208.21} & \$66.34 & \$16.29 & \$115.37 & \$73.22 & \$875.05\\
  \bottomrule
\end{tabular}
}
\end{table}

\begin{table}[t]
    \small
    \caption{Comparison of token count across different agent frameworks on three tasks.}
    \label{tab:token-sd-vd-pr}
    \setlength{\tabcolsep}{1pt}
    \resizebox{\textwidth}{!}{
    \begin{tabular}{lcccccccccc}
        \toprule
        \textbf{Task}
        & \textbf{Component}
        & \textbf{\agentorchestra} 
        & \textbf{\owl} 
        & \makecell[c]{\textbf{\seagent}\\ \textbf{(Iter-1)}} 
        & \makecell[c]{\textbf{\seagent}\\ \textbf{(Iter-2)}} 
        & \makecell[c]{\textbf{\seagent}\\ \textbf{(Iter-3)}} 
        & \textbf{\trae} 
        & \textbf{\gptswarm} 
        & \textbf{\openhands} 
        & \textbf{\sweagent}
        \\
        \midrule
        \multirow{2}{*}{\makecell[l]{Software\\ Development}} 
        & Input & 380.34 M & 42.88 M  & 440.69 M & 28.31 M & 20.54 M & 32.09 M & 287.31 K & \textbf{1.26 B} & 371.70 M \\
        & Output & 22.71 M & 3.34 M & 8.53 M & 11.15 M & 9.26 M & 23.09 M & 7.98 M & \textbf{30.54 M} & 814.04 K \\
        \midrule
        \multirow{2}{*}{\makecell[l]{Vulnerability\\ Detection}} 
        & Input & 36.30 M & 2.64 M & 216.91 K & 360.03 K & 7.98 M & 1.57 M & 13.66 K &  \textbf{192.75 M} & 4.17 M \\
        & Output & 1.60 M & 632.38 K & 248.32 K & 202.15 K & 537.87 K & 841.91 K & 394.12 K & \textbf{5.85 M} & 227.90 K \\
        \midrule
        \multirow{2}{*}{\makecell[l]{Program\\ Repair}} 
        & Input & 99.59 M & 6.72 M & 332.03 M & 447.96 M & 484.44 M & 13.28 M & 54.10 K & \textbf{663.55 M} & 495.50 M \\
        & Output & 2.16 M & 220.74 K & 524.66 K & 853.67 K & 839.23 K & \textbf{4.82 M} & 485.46 K & 4.56 M & 836.83 K \\
        \bottomrule
    \end{tabular}
    }
\end{table}

To address RQ3, we evaluate the costs of seven agents across three tasks. Comprehensive results are presented in \autoref{tab:cost-sd-vd-pr}. The first column indicates the three evaluation tasks: software development, vulnerability detection, and program repair. Columns 2-9 show the costs (USD) incurred by each of the seven agents for these three tasks, while the last column presents the total cost across all seven agents for each task. The bottom row displays the total cost for each agent across all three tasks.

The results reveal that \agentorchestra incurred the highest cost at \$370.19 across the three tasks, whereas \gptswarm solved all problems across the three tasks with merely \$16.29. Software development emerged as the most expensive task, with a total cost of \$544.90, where \agentorchestra consumed \$292.01, the highest among all agents. The seven agents collectively spent \$42.40 to complete the vulnerability detection task, with \openhands leading at \$20.91, while \seagent (Iter-1) required only \$0.29, the lowest expenditure. For the program repair task, the total cost reached \$287.75, with \agentorchestra again ranking first at \$64.05 for a single iteration. \seagent's cumulative cost across three iterations totaled \$106.91, while \gptswarm remained the most economical at just \$2.13.

Economic costs are determined by token consumption and model pricing. Since we employed the same LLM with consistent token pricing, we focus our analysis on token consumption, where the input price is \$0.56 per million tokens, and the output price is \$1.69 per million tokens. Notably, since the agent often reuses the same context across multiple interactions, the model can leverage a caching mechanism to reuse previously processed inputs. Consequently, even with a large number of input tokens, the overall cost may remain low due to the lower pricing of cached tokens (\$0.07 per million).  \autoref{tab:token-sd-vd-pr} presents the token consumption results for these agents, organized into three major rows representing the three tasks. Each major row contains two sub-rows indicating input and output token counts, respectively, with each column representing the token consumption for each agent.

As shown in \autoref{tab:token-sd-vd-pr}, despite \agentorchestra incurring the highest cost in the software development task, it did not consume the most tokens. Instead, \openhands consumed 1.26 B input tokens and 30.54 M output tokens, making it the highest consumer. \gptswarm consumed the fewest input tokens at 287.31 K, while \sweagent used only 814.04 K output tokens, the lowest among all agents. This discrepancy is attributable to the caching mechanism: although \openhands processed the highest number of input and output tokens, these largely consisted of accumulated historical data that, when fed back to the agent, did not incur proportional costs. In the vulnerability detection task, \openhands again consumed the most input tokens at 192.75 M and output tokens at 5.85 M. \gptswarm used the fewest input tokens at 13.66 K, while \sweagent consumed the least output tokens at 227.90 K. For the program repair task, \openhands consumed 663.55 M input tokens, the highest among all agents, while \trae generated the most output tokens at 4.82 M. In contrast, \gptswarm consumed merely 54.10 K input tokens, and \owl used 220.74 K output tokens.
A notable observation is that, except for \gptswarm, all agents consumed more input tokens than output tokens. For \gptswarm, this pattern is reversed, with output token consumption exceeding input token consumption.

\begin{figure}[t]
  \centering
  \includegraphics[width=0.9\linewidth]{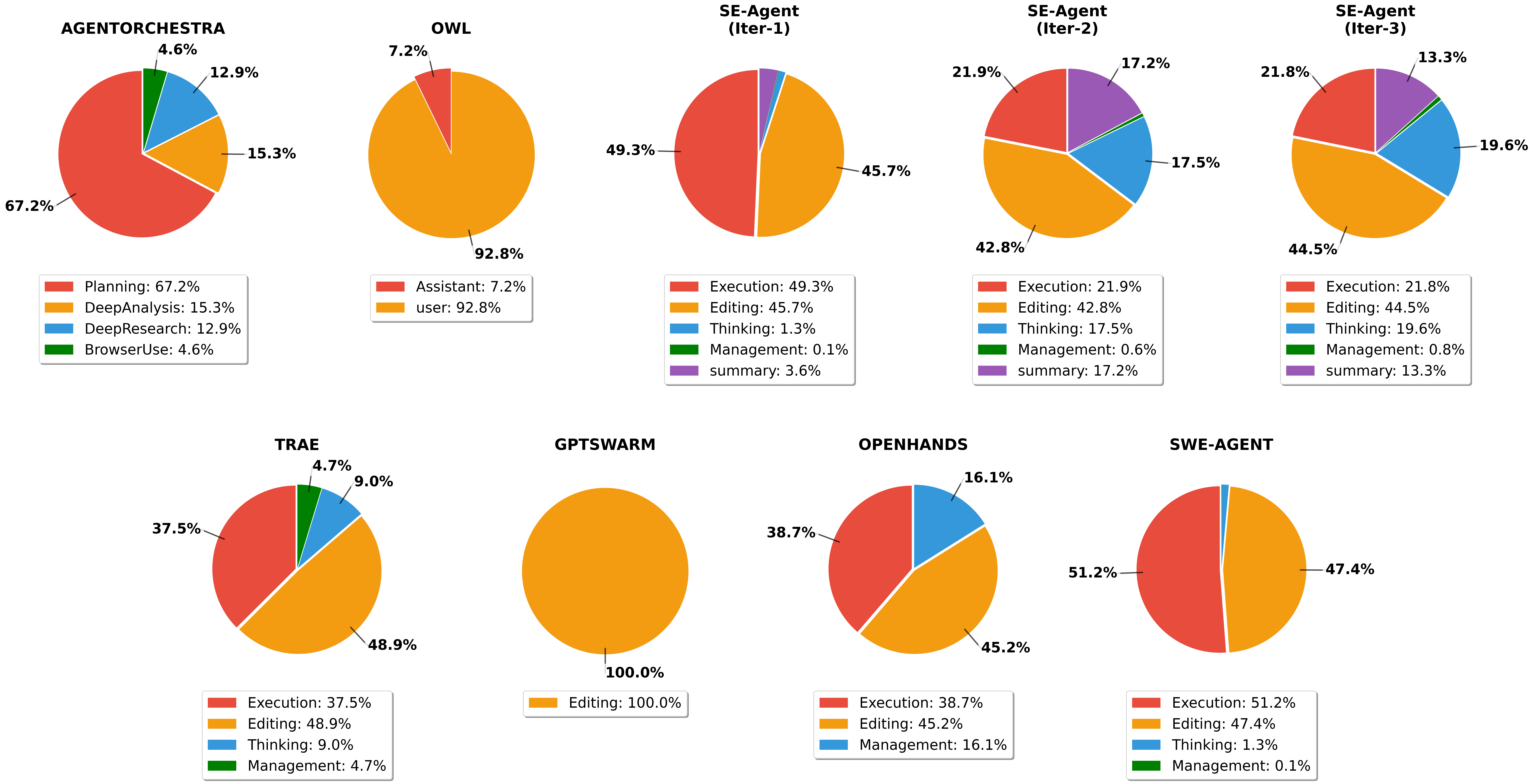}
  \vspace{-3mm}
  \caption{Breakdown of token consumption across execution stages for software development among the seven agent frameworks.}
  \vspace{-3mm}
  \label{fig:sd-pie-charts}
\end{figure}

\begin{figure}[t]
  \centering
  \includegraphics[width=0.9\linewidth]{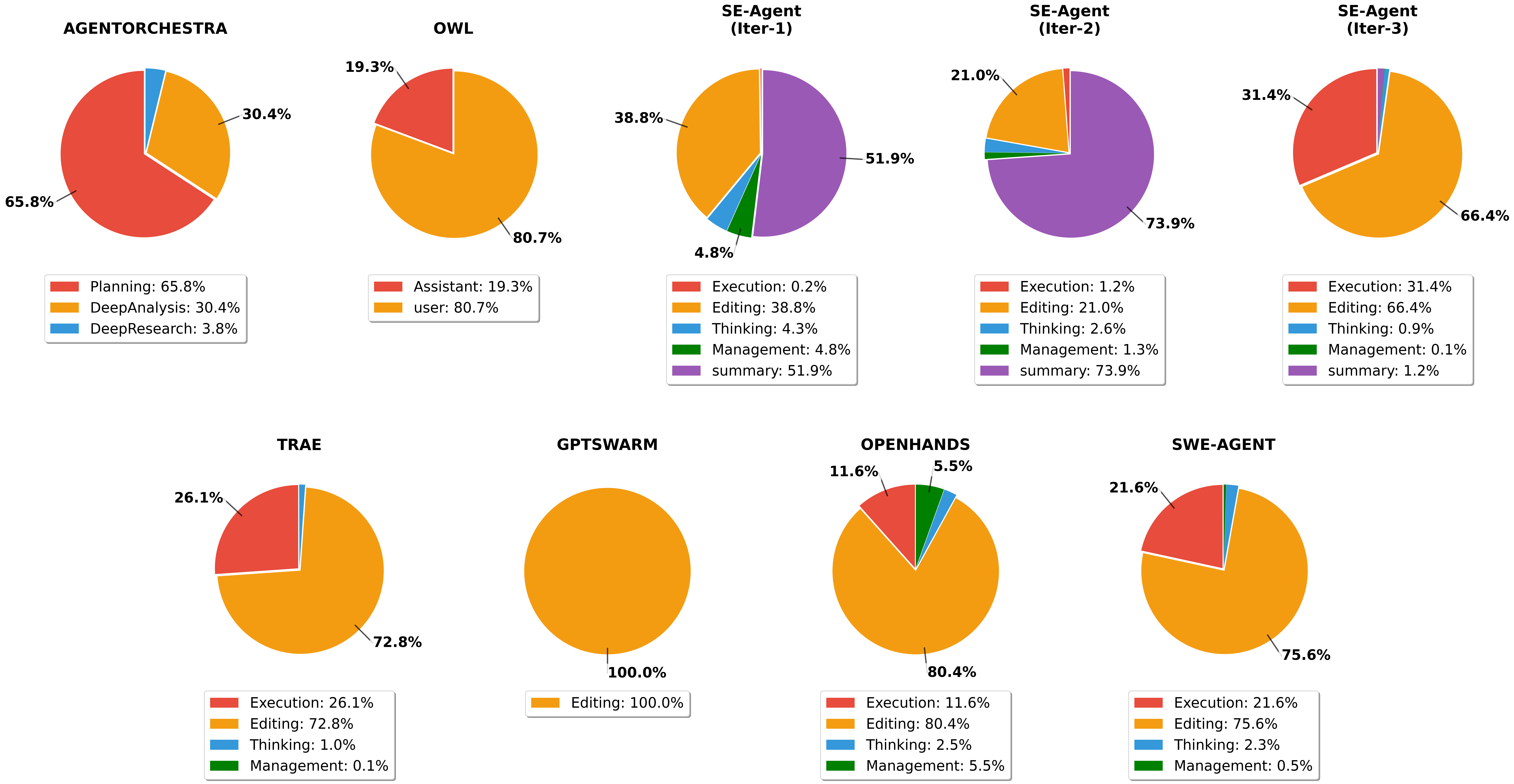}
  \vspace{-3mm}
  \caption{Breakdown of token consumption across execution stages for vulnerability detection among the seven agent frameworks.}
  \vspace{-3mm}
  \label{fig:vd-pie-charts}
\end{figure}

\begin{figure}[t]
  \centering
  \includegraphics[width=0.9\linewidth]{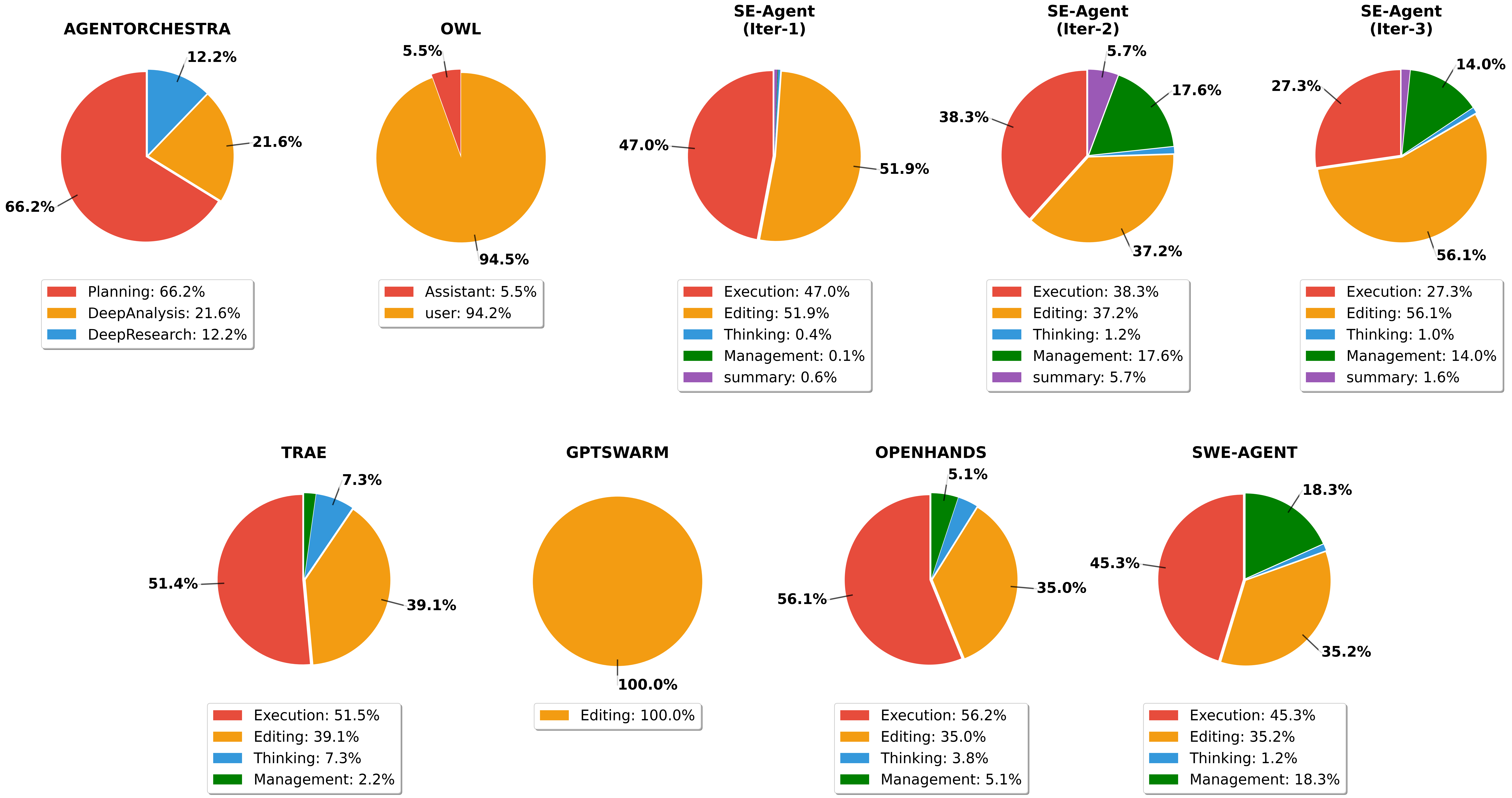}
  \vspace{-3mm}
  \caption{Breakdown of token consumption across execution stages for program repair among the seven agent frameworks.}
  \vspace{-3mm}
  \label{fig:pr-pie-charts}
\end{figure}

To further investigate how token consumption is distributed across different execution stages among the seven agent frameworks, we employ the classification framework presented in \autoref{fig:sd-pie-charts} for software development, \autoref{fig:vd-pie-charts} for vulnerability detection, and \autoref{fig:pr-pie-charts} for program repair. The first row comprises two multi-agent systems
for which we analyze token consumption at the individual sub-agent level. Given that \seagent's worker agents operate analogously to \sweagent as single agents and utilize a summary operators for post-evaluation trajectory synthesis, we apply a unified statistical methodology to \seagent's worker agents along with the four single-agent systems shown in the second row,
decomposing token consumption into four action categories: execution, editing, thinking, and management.

\vspace{1mm}
\noindent
\textbf{Software Development}
As illustrated in \autoref{fig:sd-pie-charts}, \agentorchestra employs four distinct agents for software development tasks: Planning agent, DeepAnalysis agent, DeepResearch agent, and BrowserUse agent. The token consumption is predominantly attributed to the Planning agent, which accounts for 67.2\% of the total consumption, followed by the DeepAnalysis agent at 15.3\% and the DeepResearch agent at 12.9\%, while the BrowserUse agent contributes the least at 4.6\%. \owl comprises two agents: User agent and Assistant agent, where the User agent dominates token consumption at 92.8\%, with the Assistant agent accounting for merely 7.2\%.

For \seagent in its first iteration, token consumption is primarily distributed between editing and execution, representing 49.3\% and 45.7\% of the total respectively, whereas management and thinking constitute only 0.1\% and 1.3\%. This distribution stems from \seagent's well-structured prompts with examples, and the post-evaluation summary accounts for only 3.6\% of total consumption. In subsequent iterations, the streamlined prompts afford \seagent greater flexibility, resulting in increased proportions of management and thinking activities. Nevertheless, editing and execution continue to dominate total consumption, while the reduced prompt complexity leads to elevated summary operators consumption of 17.7\% and 13.3\% in the second and third iterations respectively.

\autoref{fig:sd-pie-charts} further reveals that single-agent systems allocate the majority of token consumption to editing activities in software development tasks, specifically 48.9\% for \trae and 45.2\% for \openhands. \gptswarm exclusively utilizes editing due to its IO-agent architecture, resulting in 100\% editing consumption. Conversely, \sweagent exhibits the highest execution consumption at 51.2\% of total tokens, with editing as the second-largest category at 47.4\%. \trae and \openhands demonstrate execution consumption of 37.5\% and 38.7\% respectively. Thinking activities constitute a minimal proportion in \openhands and \sweagent, while accounting for 9.0\% in \trae. Similarly, management activities represent a negligible share in \sweagent, whereas \trae and \openhands allocate 4.7\% and 16.1\% of total consumption to management respectively.

\vspace{1mm}
\noindent
\textbf{Vulnerability Detection}
As depicted in \autoref{fig:vd-pie-charts}, \agentorchestra employs three agents for vulnerability detection tasks: Planning agent, DeepAnalysis agent, and DeepResearch agent. Token consumption is predominantly concentrated in the Planning agent at 65.8\% of the total, followed by the deep analysis agent at 30.4\% and the deep research agent at 3.8\%. For \owl, the User agent accounts for 80.7\% of total consumption, while the Assistant agent represents 19.3\%.

In \seagent's first two iterations, token consumption is overwhelmingly dominated by the summary process, constituting 51.9\% and 73.9\% of total consumption respectively, whereas the third iteration allocates only 1.2\% to summary activities. Editing operations account for 38.8\%, 21.0\%, and 66.4\% across the three iterations respectively. Execution activities demonstrate minimal presence in the first two iterations but increase substantially to 31.4\% in the third iteration, while thinking and management activities maintain negligible proportions throughout all iterations.

\autoref{fig:vd-pie-charts} further demonstrates that single-agent systems allocate the majority of token consumption to editing activities in vulnerability detection tasks, specifically 72.8\% for \trae, 80.4\% for \openhands, and 75.6\% for \sweagent. Execution represents the second-largest consumption category across these three agents, accounting for 26.1\%, 11.6\%, and 21.6\% respectively. Other activities including thinking and management constitute minimal proportions across all three agents. These three agents exhibit remarkable consistency in their token consumption patterns when addressing vulnerability detection tasks.

\vspace{1mm}
\noindent
\textbf{Program Repair}
As illustrated in \autoref{fig:pr-pie-charts}, \agentorchestra employs three agents for program repair tasks: Planning agent, DeepAnalysis agent, and DeepResearch agent. Token consumption is primarily attributed to the Planning agent at 66.2\% of the total, followed by the DeepAnalysis agent at 21.6\%, with the DeepResearch agent contributing the least at 12.2\%. \owl comprises two agents: User agent and Assistant agent, where the User agent dominates token consumption at 94.2\%, while the Assistant agent accounts for merely 5.5\%.

For \seagent in its first iteration, token consumption is predominantly distributed between editing and execution, representing 51.9\% and 47.0\% of the total respectively, whereas management and thinking constitute only 0.1\% and 0.4\%. This distribution is attributable to \seagent's well-structured prompts with examples, and the post-evaluation summary accounts for only 0.6\% of total consumption. In subsequent iterations, the streamlined prompts afford \seagent greater flexibility, resulting in increased proportions of management and thinking activities. Nevertheless, editing and execution continue to dominate total consumption, while the reduced prompt complexity leads to elevated summary operators consumption of 5.7\% and 1.6\% in the second and third iterations respectively.

\autoref{fig:pr-pie-charts} further reveals that single-agent systems allocate the majority of token consumption to execution activities in program repair tasks, specifically 51.5\% for \trae, 56.2\% for \openhands, and 45.3\% for \sweagent. Editing operations account for 39.1\%, 100.0\%, 35.0\%, and 35.2\% of total consumption for \trae, \gptswarm, \openhands, and \sweagent respectively. While thinking and management activities constitute minimal proportions in \trae and \openhands, management activities notably account for 18.3\% of total consumption in \sweagent.


\begin{tcolorbox}[size=title]
\textbf{Answer to RQ3.} The key takeaways are as follows:
\begin{itemize}
    \item \textbf{Monetary Cost:} Among the three tasks, software development incurs the highest cost, while vulnerability detection requires the lowest. Of the seven agents, \agentorchestra is the most expensive, whereas \gptswarm is the most cost-efficient.
    \item \textbf{Token Usage:} \openhands consumes the most tokens across all three tasks but does not incur the highest cost due to input token caching. \gptswarm consumes the fewest tokens and is the only agent where input tokens are fewer than output tokens, while all other agents exhibit higher input token consumption.
    \item \textbf{Consumption Breakdown:} Agent systems exhibit similar patterns across tasks. For example,in \agentorchestra, the Planning agent dominates. \seagent shows varying distributions due to prompt modifications. Single-agent systems focus consumption on execution and editing actions.

\end{itemize}
\end{tcolorbox}

\section{Discussion}


After independently analyzing the effectiveness and efficiency of the evaluated agent frameworks, we now explore their interrelationships, with a particular focus on the intuitive notion that more complex reasoning and execution processes often lead to improved performance outcomes. We then address the potential threats to validity in our study to provide a balanced understanding of the results and their generalizability.

\subsection{More steps, better effectiveness?}

\begin{figure}[t]
  \centering
  \includegraphics[width=0.8\linewidth]{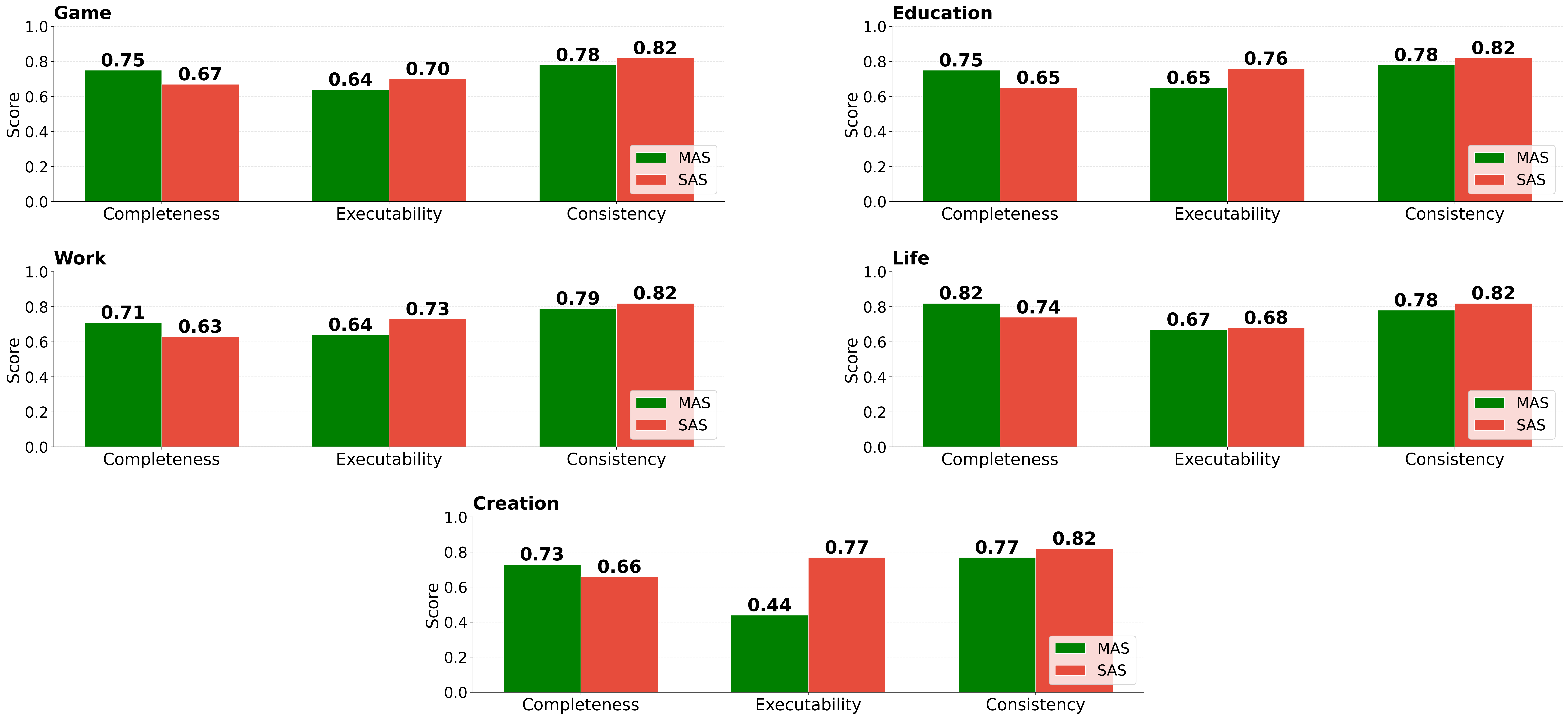}
  \vspace{-3mm}
  \caption{multi-agent systems vs single-agent systems in Software Development.}
  \vspace{-3mm}
  \label{fig:sd-mas-sas}
\end{figure}

To facilitate a more comprehensive discussion of this issue, we categorize the seven agents into two groups based on the findings of RQ2: multi-agent systems (Multi-Agent system) and single-agent systems (Single-Agent system). This classification is motivated by the observation that multi-agent systems exhibit substantially longer average steps across all three tasks compared to their single-agent counterparts, with the disparity being particularly pronounced in the vulnerability detection and program repair task, where multi-agent systems require nearly twice the number of steps. Building upon the results from RQ1, we further conduct a statistical analysis to determine whether multi-agent systems leverage their extended execution trajectories to achieve superior effectiveness compared to single-agent systems across the three tasks.

\vspace{1mm}
\noindent
\textbf{Software Development.}
To systematically compare the differences between multi-agent systems and single-agent systems  in software development tasks, we computed the average completeness, executability, and consistency of code projects generated by multi-agent systems and single-agent systems agents across five software development categories. The comparative results are illustrated in \autoref{fig:sd-mas-sas}.
The analysis reveals nuanced performance trade-offs between the two paradigms across three evaluation dimensions. In terms of completeness, multi-agent systems demonstrates superior performance across all five software categories, consistently outperforming single-agent systems metrics. This suggests that Multi-Agent frameworks are more effective at generating comprehensive solutions that address all specified requirements.

Regarding executability, the generated code by single-agent systems exhibits a substantial advantage, significantly surpassing multi-agent systems performance. This indicates that single-agent systems produce more syntactically correct and immediately executable code. In the consistency evaluation, single-agent systems marginally outperforms multi-agent systems, suggesting slightly better adherence to coding standards and internal coherence. \openhands demonstrates the best quality of 0.47. Overall, single-agent systems demonstrates greater effectiveness in software development, particularly excelling in code executability and consistency. When combined with the conclusions of RQ2, it is evident that single-agent systems holds a slight advantage over multi-agent systems.

\begin{figure}[t]
  \centering
  \includegraphics[width=\linewidth]{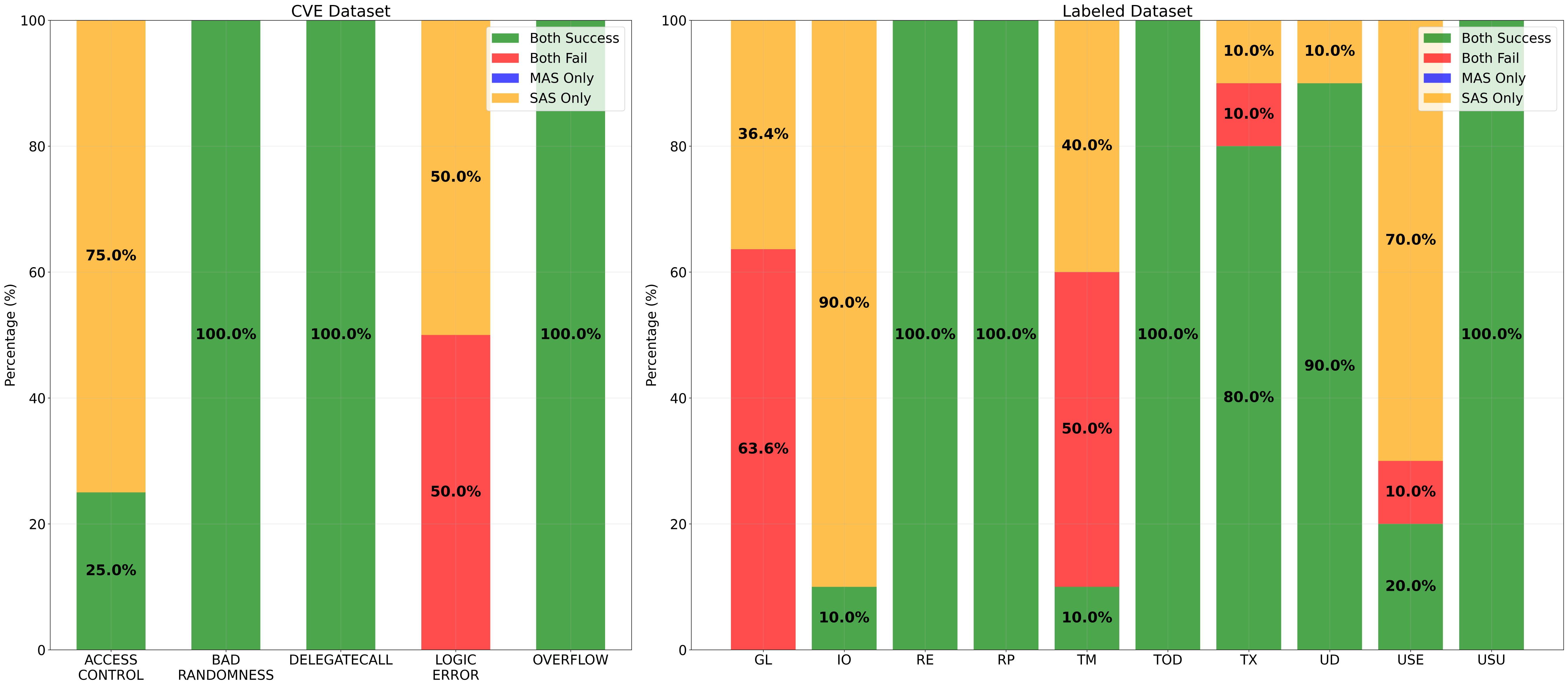}
  \vspace{-3mm}
  \caption{multi-agent systems vs single-agent systems in Vulnerability Detection.}
  \vspace{-3mm}
  \label{fig:vd-mas-sas}
\end{figure}

\vspace{1mm}
\noindent
\textbf{Vulnerability Detection.}
We focused on two critical subsets: (1) the union of successfully completed tasks by both system types, and (2) the intersection of tasks that neither system could complete. 
To compare the efficiency of multi-agent systems and single agent systems in smart contract vulnerability detection, we categorized the detection results into multi-agent systems and single-agent systems groups and conducted a comparative analysis across 10 vulnerability types of common dataset and 5 vulnerability types of CVE. This approach provides a more intuitive visualization of detection accuracy, as illustrated in \autoref{fig:vd-mas-sas}. The results demonstrate that multi-agent systems and single-agent systems exhibit highly consistent detection effectiveness across nine vulnerability types, including Bad Randomness and RE. However, single-agent systems outperforms multi-agent systems in detecting six specific vulnerability types: Access Control, Logic Error, GL, IO, TM, and USE. Notably, no vulnerability types were exclusively detected by multi-agent systems. According to the results of RQ2, the number of steps in multi-agent systems is higher than that in single-agent systems, therefore single-agent systems demonstrates an advantage in this task.

\begin{figure}[t]
  \centering
  \includegraphics[width=\linewidth]{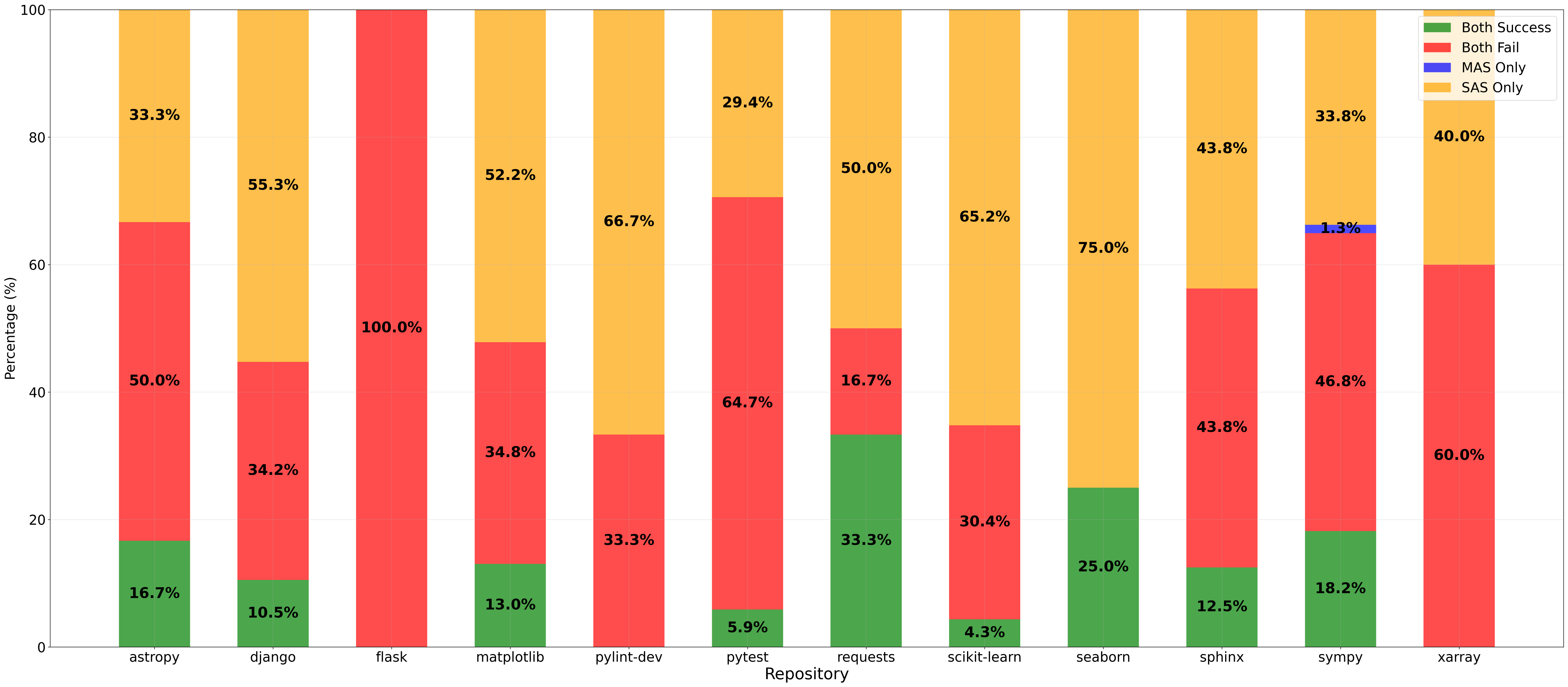}
  \vspace{-3mm}
  \caption{multi-agent systems vs single-agent systems in Program Repair.}
  \vspace{-3mm}
  \label{fig:pr-mas-sys}
\end{figure}

\vspace{1mm}
\noindent
\textbf{Program Repair.}
To compare the problem-solving efficiency between multi-agent systems and single-agent systems, we conducted a controlled analysis by selecting two distinct subsets: (1) the union of successfully repaired issues by both system types, and (2) the intersection of issues that neither system could repair. The statistical results are presented in \autoref{fig:pr-mas-sys}.
Similarly, we partition the experimental results into multi-agent systems and single-agent systems categories. Our analysis reveals that single-agent systems demonstrate superior repair effectiveness across all 12 repositories compared to multi-agent systems. When synthesized with the empirical evidence from RQ2, these findings substantiate that single-agent systems possesses a pronounced advantage in program repair tasks.

\vspace{1mm}
\noindent
\textbf{Reason Analysis.} Based on the statistical analysis across the three aforementioned tasks, it is evident that single-agent system consistently demonstrates superior performance compared to multi-agent system in software engineering applications. To investigate this phenomenon, we conducted a comprehensive analysis of the underlying mechanisms.

In multi-agent system, multiple agents typically collaborate by invoking shared tools or utilizing specialized agent-specific tools, with a planning agent orchestrating the problem-solving workflow. Conversely, single-agent system architectures employ a single agent equipped with tools and follow a predefined workflow tailored to each specific task. However, the proliferation of agents in multi-agent system introduces substantial interaction overhead between the planning agent and downstream specialized agents. This results in information overload for the planning agent, leading to decision-making errors characterized by overthinking and erroneous outputs. Furthermore, inter-agent hallucinations significantly impact overall accuracy. As demonstrated in RQ2, multi-agent system exhibits higher correction attempts and correction rates compared to single-agent system. This phenomenon is attributable to two primary factors: (1) the increased number of agents generates excessive input tokens that exceed the LLM's maximum context length, resulting in information loss, and (2) inherent LLM hallucinations. In contrast, single-agent system operates with a single agent, generating minimal context that remains within the LLM's long-context constraints, thereby enabling comprehensive access to historical information and better control over FPs and hallucinations. These factors collectively contribute to multi-agent system underperforming single-agent system across all three tasks.

Theoretically, the presence of planning mechanisms in multi-agent system should confer superior generalization capabilities; however, our empirical findings reveal the contrary. As previously noted, frameworks such as \agentorchestra and \owl fail to generate correctly formatted \codeff{diff} patches, resulting in diminished repair success rates in program repair tasks. This indicates that multi-agent collaboration is susceptible to propagating identical errors, where agents may fail to detect mistakes in outputs generated by other agents. In contrast, specialized repair agents like \sweagent are equipped with dedicated patch tools that circumvent format-related failures. Similarly, \gptswarm's lack of specialized tools leads to the generation of malformed \codeff{diff} patches, consequently reducing repair rates.

Notably, \seagent deviates from conventional multi-agent collaborative generation paradigms. Instead, it employs a single agent for task execution, followed by a summary agent that synthesizes trajectory information. In subsequent iterations, the worker agent learns from successful experiences or avoids failed trajectories—a process we characterize as vertical iteration. Our analysis reveals that \seagent's performance improves progressively across three iterations. For instance, in program repair tasks, the initial repair success rate of 47.33\% improved to 53.00\% after the first summarization iteration, and ultimately reached 53.67\% after two summarization rounds and three complete iterations.

Another noteworthy approach is \gptswarm's group execution strategy, which partitions instances into predetermined groups within a single testing round. Upon completing each group, the agent synthesizes lessons learned from both successful and failed cases, providing experiential guidance for subsequent groups. This methodology proves particularly advantageous in vulnerability detection tasks, making \gptswarm the top-performing agent in accuracy.

These findings collectively demonstrate that in software engineering lifecycle tasks, incorporating specialized tools yields superior results compared to adding dedicated agents, facilitating rapid generalization and adaptation to novel tasks. Moreover, both vertical and horizontal summarization of historical trajectories during task execution effectively enhance agent performance, providing a training-free approach to performance augmentation.

\subsection{Threats to Validity}
\textbf{Internal.} Several factors may influence the internal validity of our study.  
(1) Regarding the selection of agent frameworks, proprietary task-specific agents may perform optimally on particular tasks; however, they lack generality, which is inconsistent with the current trend toward developing AGI. Therefore, we focus on general-purpose agent frameworks, which are more cutting-edge, have broader applicability, and allow fair comparisons across multiple tasks.  
(2) Concerning dataset selection, an ideal general-purpose agent would handle the full software engineering lifecycle, from requirements to reliable code. However, evaluating each stage of the lifecycle is challenging and resource-intensive, and no single benchmark currently covers the entire process. To ensure objective evaluation, we concentrate on key code-centric tasks and adopt existing benchmarks for each task.  
(3) Regarding the choice of underlying LLMs, we acknowledge that different base models can lead to variations in agent performance. Considering the considerable economic overhead of agents, we employed DeepSeek-V3.1 as the underlying LLM consistently across all agents and tasks. This controlled approach allows for comparable results. In future work, when resources permit, it would be worthwhile to explore different base models and potential model selection optimizations.

\vspace{1mm}
\noindent
\textbf{External.} 
The primary external validity threats stem from implementation discrepancies between the published frameworks and their released codebases. For instance, certain auxiliary components described in \trae’s design are not fully reflected in the official example code. We retained the functional integrity of the released implementation and fixed random seeds to ensure consistent comparison. Although the results may not fully represent the idealized design, they faithfully reflect the framework’s practical behavior in real-world use. Additionally, some frameworks adopt simplified configurations in their released code for code-related tasks. We therefore explicitly reported the actual agent and role settings used in our experiments to avoid overinterpreting their collaborative capabilities.
\section{Related Work}

As discussed in \S~\ref{subsec:dataset}, the use of intelligent agents in code-centric software engineering has recently attracted increasing research attention. Numerous frameworks have been introduced to automate a range of tasks, including software development, vulnerability detection, and program repair. However, to the best of our knowledge, this work is the first to conduct a systematic empirical comparison of contemporary agent-based approaches, aiming to provide a deeper understanding of their actual capabilities and limitations across diverse software engineering scenarios.

\vspace{1mm}
\textbf{Agents for Software Engineering.}
In recent years, a growing number of versatile agent frameworks have emerged and been applied to increasingly complex scientific and productivity tasks~\cite{agentsurvey2025}. As software engineering represents a core pillar of productivity, enhancing its automation through agent technologies holds substantial promise for improving development efficiency and reliability. Accordingly, many studies have explored the application of intelligent agents to code-centric SE tasks, as discussed in \S~\ref{subsec:agent}.
Among them, some agent frameworks are designed for specific tasks, such as \experepair \cite{experepair} and \semagent \cite{semagent}, which focus exclusively on program repair.
In contrast, general-purpose agent frameworks that are capable of handling diverse complex tasks exhibit stronger generalization and better alignment with real-world software engineering workflows. For instance, to support complex software engineering environments, \sweagent \cite{sweagent} enhances repository-scale operations with a customized agent–computer interface that enables autonomous file manipulation, repository navigation, and test execution. Likewise, \trae \cite{trae} provides a research-friendly command-line interface for orchestrating real-world development workflows in a extensible manner. To align agent behavior more closely with human development practices, \openhands \cite{openhands} equips agents with human-like operational abilities, such as interacting with terminals and browsing the web. To better orchestrate agent behaviors, \agentorchestra \cite{agentorchestra} employs hierarchical task decomposition through a top-level planner directing multiple specialized workers, while \owl \cite{owl} structures multi-agent collaboration following a workforce-style paradigm aimed at large-scale productivity. \gptswarm \cite{gptswarm} further advances scalable coordination by leveraging graph-based composition and self-organized swarm intelligence. In addition, to optimize the reasoning trajectory, \seagent \cite{seagent} proposes a self-evolving multi-step reasoning paradigm that refines solutions through strategy diversification and trajectory optimization.
Overall, these developments highlight the growing convergence between agent technologies and software engineering, underscoring the need for a systematic evaluation of their actual capabilities across diverse software engineering tasks.

\vspace{1mm}
\textbf{Benchmarks for Software Engineering Agents.}
Although intelligent agents have demonstrated the potential to handle multiple phases of the software engineering lifecycle in an end-to-end manner, the community still lacks a comprehensive benchmark capable of holistically evaluating their capabilities across different development stages. Existing evaluations remain fragmented and rely on task-specific datasets.

For software development tasks, widely used benchmarks include HumanEval~\cite{humaneval} and MBPP~\cite{mbpp}, which focus on single-function code generation and contain only a few hundred tasks. More complex datasets requiring multi-file development have emerged, such as SRDD~\cite{srdd}, CAASD~\cite{caasd}, SoftwareDev~\cite{metagpt}, and ProjectDev~\cite{projectdev}. With over a thousand tasks, SRDD achieves broader task coverage, while the other datasets include only dozens and therefore fall short in representing realistic, large-scale development workflows.

For vulnerability detection, curated datasets such as  SySeVR~\cite{sysevr} for C/C++, CWE-Bench-Java~\cite{cwebenchjava} for Java, and BigBench~\cite{bigbench} for Python enable the evaluation of LLM-based vulnerability prediction. However, these datasets primarily measure static detection performance while overlooking the interactive reasoning and environment orchestration capabilities expected from agents. To the best of our knowledge, LLM-SmartAudit \cite{llmsmartaudit} is the first work that introduces an agent-based approach for vulnerability detection, and its dataset thus serves as an important reference for evaluating existing agent capabilities in this domain.

For vulnerability repair, classical benchmarks like Defects4J~\cite{defects4j}, BugsInPy~\cite{bugsinpy}, HumanEval-Java~\cite{humanevaljava}, and QuixBugs~\cite{quixbugs} contain small-scale defects localized in specific functions or modules. They are suitable for evaluating traditional automated repair techniques or single LLM capabilities, but insufficient for assessing multi-step repair behaviors in realistic engineering environments.
To better align with real-world workflows, recent benchmarks construct repair tasks directly from GitHub issues and their associated repositories, such as SWE-bench~\cite{swebench}, SWE-bench Lite~\cite{swebenchlite}, SWE-bench Lite-S~\cite{swebenchlites}, and SWE-bench Verified~\cite{swebenchverified}, which are widely adopted for evaluating agent-based program repair.

Overall, no single benchmark can comprehensively evaluate the full spectrum of capabilities exhibited by software engineering agents. A unified evaluation framework that integrates diverse datasets across different task categories is needed to fully assess their robustness and generalizability throughout the entire software engineering life-cycle.


\vspace{1mm}
\textbf{Empirical Evaluation of Software Engineering Agents.}
The recent surge in intelligent software engineering agents has drawn increasing attention from the research community, prompting systematic empirical studies of their capabilities. Researchers have examined agent behavior from multiple perspectives, including architectural design, reasoning processes, and task-specific effectiveness.

Regarding architectural design, studies focus on the intrinsic characteristics of agent systems, such as single-agent versus multi-agent configurations. Gao \etal~\cite{gao2025single} conducted an extensive comparison across 15 software engineering tasks, showing that multi-agent systems generally perform better on complex tasks due to long-horizon context tracking and role-specific error correction, whereas single-agent systems often achieve higher efficiency on simpler tasks. They further proposed a hybrid paradigm, cascading requests between multi-agent systems and single-agent systems, to balance accuracy and deployment cost.

In terms of reasoning processes, empirical analyses have investigated agent decision-making workflows. Ceka \etal~\cite{ceka2025understanding} present the first systematic study of software engineering agents through the lens of execution traces. They classify decision pathways, identify core components such as bug localization, patch generation, and test generation, and analyze how reasoning about error context influences successful outcomes. These insights illuminate the internal mechanisms that drive agent performance, independent of specific task instances.

Moreover, task-specific evaluations provide complementary evidence of agent effectiveness. Meng \etal~\cite{meng2025empiricalstudyllmbasedagents} conducted fine-grained empirical studies of LLM-based agents for automated bug fixing, assessing fault localization accuracy at file- and line-levels, bug reproduction capabilities, and overall repair performance. Their findings highlight that improving agent effectiveness requires advances in both the underlying LLM models and the design of agent workflows.

Unlike previous studies, which often focus on specific tasks or architectural configurations, our work provides a systematic evaluation of general-purpose software engineering agents across a range of code-centric tasks. We assess not only their effectiveness but also their efficiency and deployment cost, offering a holistic perspective on agent performance throughout the software engineering life-cycle. This comprehensive approach allows us to identify strengths and limitations of current frameworks and informs the design of more capable and practical agent systems.



\section{Conclusion}

This study presents a comprehensive empirical evaluation of seven general-purpose agent frameworks across three critical code-centric software engineering tasks: software development, vulnerability detection, and program repair. By systematically examining agent performance from the perspectives of effectiveness, efficiency, and overhead, we reveal important insights into the current capabilities and limitations of intelligent agents in real-world software engineering scenarios.
Our study reveals important insights into agent frameworks from three key angles. In terms of effectiveness, agents show moderate success: \openhands balances code quality well in software development, \gptswarm excels in vulnerability detection accuracy, and program repair remains challenging with only some agents fixing about half of the issues. Efficiency analysis indicates that \seagent(Iter-3) requires the most steps across all experiments. \agentorchestra has the longest correction sequences, while \openhands ranks second, demonstrating stable and convergent behavior. Regarding overhead, software development tasks are the most costly. The token consumption breakdown reveals multi-agent frameworks dominated by planning stages, while single-agent systems concentrate costs on execution and editing activities. These findings collectively highlight the trade-offs and potentials of current frameworks, guiding future improvements in building more capable and resource-aware software engineering agents.

\bibliographystyle{ACM-Reference-Format}
\bibliography{manu}


\begin{thebibliography}{47}


\ifx \showCODEN    \undefined \def \showCODEN     #1{\unskip}     \fi
\ifx \showISBNx    \undefined \def \showISBNx     #1{\unskip}     \fi
\ifx \showISBNxiii \undefined \def \showISBNxiii  #1{\unskip}     \fi
\ifx \showISSN     \undefined \def \showISSN      #1{\unskip}     \fi
\ifx \showLCCN     \undefined \def \showLCCN      #1{\unskip}     \fi
\ifx \shownote     \undefined \def \shownote      #1{#1}          \fi
\ifx \showarticletitle \undefined \def \showarticletitle #1{#1}   \fi
\ifx \showURL      \undefined \def \showURL       {\relax}        \fi
\providecommand\bibfield[2]{#2}
\providecommand\bibinfo[2]{#2}
\providecommand\natexlab[1]{#1}
\providecommand\showeprint[2][]{arXiv:#2}

\bibitem[swe(2024a)]%
        {swebenchlite}
 \bibinfo{year}{2024}\natexlab{a}.
\newblock \bibinfo{booktitle}{\emph{SWE-bench Lite}}.
\newblock
\urldef\tempurl%
\url{https://www.swebench.com/lite.html}
\showURL{%
\tempurl}


\bibitem[swe(2024b)]%
        {swebenchverified}
 \bibinfo{year}{2024}\natexlab{b}.
\newblock \bibinfo{booktitle}{\emph{SWE-bench Verified}}.
\newblock
\urldef\tempurl%
\url{https://openai.com/index/introducing-swe-bench-verified/}
\showURL{%
\tempurl}


\bibitem[ant(2025)]%
        {anthropic-agent}
 \bibinfo{year}{2025}\natexlab{}.
\newblock \bibinfo{booktitle}{\emph{Building effective agents.}}
\newblock
\urldef\tempurl%
\url{https://www.anthropic.com/engineering/building-effective-agents}
\showURL{%
\tempurl}


\bibitem[goo(2025)]%
        {google-agent}
 \bibinfo{year}{2025}\natexlab{}.
\newblock \bibinfo{booktitle}{\emph{Google's Whitepaper on Agents.}}
\newblock
\urldef\tempurl%
\url{https://drive.google.com/file/d/1oEjiRCTbd54aSdB_eEe3UShxLBWK9xkt/view?pli=1}
\showURL{%
\tempurl}


\bibitem[Aggarwal(2005)]%
        {aggarwal2005software}
\bibfield{author}{\bibinfo{person}{Krishan~Kumar Aggarwal}.} \bibinfo{year}{2005}\natexlab{}.
\newblock \bibinfo{booktitle}{\emph{Software engineering}}.
\newblock \bibinfo{publisher}{New Age International}.
\newblock


\bibitem[Austin et~al\mbox{.}(2021)]%
        {mbpp}
\bibfield{author}{\bibinfo{person}{Jacob Austin}, \bibinfo{person}{Augustus Odena}, \bibinfo{person}{Maxwell Nye}, \bibinfo{person}{Maarten Bosma}, \bibinfo{person}{Henryk Michalewski}, \bibinfo{person}{David Dohan}, \bibinfo{person}{Ellen Jiang}, \bibinfo{person}{Carrie Cai}, \bibinfo{person}{Michael Terry}, \bibinfo{person}{Quoc Le}, {et~al\mbox{.}}} \bibinfo{year}{2021}\natexlab{}.
\newblock \showarticletitle{Program synthesis with large language models}.
\newblock \bibinfo{journal}{\emph{arXiv preprint arXiv:2108.07732}} (\bibinfo{year}{2021}).
\newblock


\bibitem[Basili(1989)]%
        {basili1989software}
\bibfield{author}{\bibinfo{person}{Victor~R Basili}.} \bibinfo{year}{1989}\natexlab{}.
\newblock \showarticletitle{Software development: A paradigm for the future}. In \bibinfo{booktitle}{\emph{[1989] Proceedings of the Thirteenth Annual International Computer Software \& Applications Conference}}. IEEE, \bibinfo{pages}{471--485}.
\newblock


\bibitem[Bessey et~al\mbox{.}(2010)]%
        {bessey2010afew}
\bibfield{author}{\bibinfo{person}{Al Bessey}, \bibinfo{person}{Ken Block}, \bibinfo{person}{Ben Chelf}, \bibinfo{person}{Andy Chou}, \bibinfo{person}{Bryan Fulton}, \bibinfo{person}{Seth Hallem}, \bibinfo{person}{Charles Henri-Gros}, \bibinfo{person}{Asya Kamsky}, \bibinfo{person}{Scott McPeak}, {and} \bibinfo{person}{Dawson Engler}.} \bibinfo{year}{2010}\natexlab{}.
\newblock \showarticletitle{A few billion lines of code later: using static analysis to find bugs in the real world}.
\newblock \bibinfo{journal}{\emph{Commun. ACM}} \bibinfo{volume}{53}, \bibinfo{number}{2} (\bibinfo{date}{Feb.} \bibinfo{year}{2010}), \bibinfo{pages}{66–75}.
\newblock
\showISSN{0001-0782}
\href{https://doi.org/10.1145/1646353.1646374}{doi:\nolinkurl{10.1145/1646353.1646374}}


\bibitem[Ceka et~al\mbox{.}(2025)]%
        {ceka2025understanding}
\bibfield{author}{\bibinfo{person}{Ira Ceka}, \bibinfo{person}{Saurabh Pujar}, \bibinfo{person}{Shyam Ramji}, \bibinfo{person}{Luca Buratti}, \bibinfo{person}{Gail Kaiser}, {and} \bibinfo{person}{Baishakhi Ray}.} \bibinfo{year}{2025}\natexlab{}.
\newblock \showarticletitle{Understanding Software Engineering Agents Through the Lens of Traceability: An Empirical Study}.
\newblock \bibinfo{journal}{\emph{arXiv preprint arXiv:2506.08311}} (\bibinfo{year}{2025}).
\newblock


\bibitem[Chakraborty et~al\mbox{.}(2021)]%
        {chakraborty2021deep}
\bibfield{author}{\bibinfo{person}{Saikat Chakraborty}, \bibinfo{person}{Rahul Krishna}, \bibinfo{person}{Yangruibo Ding}, {and} \bibinfo{person}{Baishakhi Ray}.} \bibinfo{year}{2021}\natexlab{}.
\newblock \showarticletitle{Deep learning based vulnerability detection: Are we there yet?}
\newblock \bibinfo{journal}{\emph{IEEE Transactions on Software Engineering}} \bibinfo{volume}{48}, \bibinfo{number}{9} (\bibinfo{year}{2021}), \bibinfo{pages}{3280--3296}.
\newblock


\bibitem[Chen et~al\mbox{.}(2021a)]%
        {humaneval}
\bibfield{author}{\bibinfo{person}{Mark Chen}, \bibinfo{person}{Jerry Tworek}, \bibinfo{person}{Heewoo Jun}, \bibinfo{person}{Qiming Yuan}, \bibinfo{person}{Henrique Ponde De~Oliveira Pinto}, \bibinfo{person}{Jared Kaplan}, \bibinfo{person}{Harri Edwards}, \bibinfo{person}{Yuri Burda}, \bibinfo{person}{Nicholas Joseph}, \bibinfo{person}{Greg Brockman}, {et~al\mbox{.}}} \bibinfo{year}{2021}\natexlab{a}.
\newblock \showarticletitle{Evaluating large language models trained on code}.
\newblock \bibinfo{journal}{\emph{arXiv preprint arXiv:2107.03374}} (\bibinfo{year}{2021}).
\newblock


\bibitem[Chen et~al\mbox{.}(2021b)]%
        {humanevaljava}
\bibfield{author}{\bibinfo{person}{Mark Chen}, \bibinfo{person}{Jerry Tworek}, \bibinfo{person}{Heewoo Jun}, \bibinfo{person}{Qiming Yuan}, \bibinfo{person}{Henrique Ponde De~Oliveira Pinto}, \bibinfo{person}{Jared Kaplan}, \bibinfo{person}{Harri Edwards}, \bibinfo{person}{Yuri Burda}, \bibinfo{person}{Nicholas Joseph}, \bibinfo{person}{Greg Brockman}, {et~al\mbox{.}}} \bibinfo{year}{2021}\natexlab{b}.
\newblock \showarticletitle{Evaluating large language models trained on code}.
\newblock \bibinfo{journal}{\emph{arXiv preprint arXiv:2107.03374}} (\bibinfo{year}{2021}).
\newblock


\bibitem[DeepSeek-AI(2024)]%
        {deepseekai2024deepseekv3technicalreport}
\bibfield{author}{\bibinfo{person}{DeepSeek-AI}.} \bibinfo{year}{2024}\natexlab{}.
\newblock \bibinfo{title}{DeepSeek-V3 Technical Report}.
\newblock
\showeprint[arxiv]{2412.19437}~[cs.CL]
\urldef\tempurl%
\url{https://arxiv.org/abs/2412.19437}
\showURL{%
\tempurl}


\bibitem[Gao et~al\mbox{.}(2025a)]%
        {gao2025survey}
\bibfield{author}{\bibinfo{person}{Huan-ang Gao}, \bibinfo{person}{Jiayi Geng}, \bibinfo{person}{Wenyue Hua}, \bibinfo{person}{Mengkang Hu}, \bibinfo{person}{Xinzhe Juan}, \bibinfo{person}{Hongzhang Liu}, \bibinfo{person}{Shilong Liu}, \bibinfo{person}{Jiahao Qiu}, \bibinfo{person}{Xuan Qi}, \bibinfo{person}{Yiran Wu}, {et~al\mbox{.}}} \bibinfo{year}{2025}\natexlab{a}.
\newblock \showarticletitle{A survey of self-evolving agents: On path to artificial super intelligence}.
\newblock \bibinfo{journal}{\emph{arXiv preprint arXiv:2507.21046}} (\bibinfo{year}{2025}).
\newblock


\bibitem[Gao et~al\mbox{.}(2025b)]%
        {gao2025single}
\bibfield{author}{\bibinfo{person}{Mingyan Gao}, \bibinfo{person}{Yanzi Li}, \bibinfo{person}{Banruo Liu}, \bibinfo{person}{Yifan Yu}, \bibinfo{person}{Phillip Wang}, \bibinfo{person}{Ching-Yu Lin}, {and} \bibinfo{person}{Fan Lai}.} \bibinfo{year}{2025}\natexlab{b}.
\newblock \showarticletitle{Single-agent or Multi-agent Systems? Why Not Both?}
\newblock \bibinfo{journal}{\emph{arXiv preprint arXiv:2505.18286}} (\bibinfo{year}{2025}).
\newblock


\bibitem[Hong et~al\mbox{.}(2023)]%
        {metagpt}
\bibfield{author}{\bibinfo{person}{Sirui Hong}, \bibinfo{person}{Mingchen Zhuge}, \bibinfo{person}{Jonathan Chen}, \bibinfo{person}{Xiawu Zheng}, \bibinfo{person}{Yuheng Cheng}, \bibinfo{person}{Ceyao Zhang}, \bibinfo{person}{Jinlin Wang}, \bibinfo{person}{Zili Wang}, \bibinfo{person}{Steven Ka~Shing Yau}, \bibinfo{person}{Zi~Hen Lin}, \bibinfo{person}{Liyang Zhou}, \bibinfo{person}{Chenyu Ran}, \bibinfo{person}{Lingfeng Xiao}, \bibinfo{person}{Chenglin Wu}, {and} \bibinfo{person}{J{\"u}rgen Schmidhuber}.} \bibinfo{year}{2023}\natexlab{}.
\newblock \showarticletitle{MetaGPT: Meta Programming for A Multi-Agent Collaborative Framework}. In \bibinfo{booktitle}{\emph{International Conference on Learning Representations}}.
\newblock
\urldef\tempurl%
\url{https://api.semanticscholar.org/CorpusID:265301950}
\showURL{%
\tempurl}


\bibitem[Hu et~al\mbox{.}(2025)]%
        {owl}
\bibfield{author}{\bibinfo{person}{Mengkang Hu}, \bibinfo{person}{Yuhang Zhou}, \bibinfo{person}{Wendong Fan}, \bibinfo{person}{Yuzhou Nie}, \bibinfo{person}{Bowei Xia}, \bibinfo{person}{Tao Sun}, \bibinfo{person}{Ziyu Ye}, \bibinfo{person}{Zhaoxuan Jin}, \bibinfo{person}{Yingru Li}, \bibinfo{person}{Qiguang Chen}, \bibinfo{person}{Zeyu Zhang}, \bibinfo{person}{Yifeng Wang}, \bibinfo{person}{Qianshuo Ye}, \bibinfo{person}{Bernard Ghanem}, \bibinfo{person}{Ping Luo}, {and} \bibinfo{person}{Guohao Li}.} \bibinfo{year}{2025}\natexlab{}.
\newblock \bibinfo{title}{OWL: Optimized Workforce Learning for General Multi-Agent Assistance in Real-World Task Automation}.
\newblock
\showeprint[arxiv]{2505.23885}~[cs.AI]
\urldef\tempurl%
\url{https://arxiv.org/abs/2505.23885}
\showURL{%
\tempurl}


\bibitem[JACKSON et~al\mbox{.}(2025)]%
        {jackson2025impact}
\bibfield{author}{\bibinfo{person}{VICTORIA JACKSON}, \bibinfo{person}{BOGDAN VASILESCU}, \bibinfo{person}{DANIEL RUSSO}, \bibinfo{person}{PAUL RALPH}, \bibinfo{person}{RAFAEL PRIKLADNICKI}, \bibinfo{person}{MALIHEH IZADI}, \bibinfo{person}{SARAH D’ANGELO}, \bibinfo{person}{SARAH INMAN}, \bibinfo{person}{ANIELLE ANDRADE}, {and} \bibinfo{person}{ANDR{\'E} VAN DER~HOEK}.} \bibinfo{year}{2025}\natexlab{}.
\newblock \showarticletitle{The Impact of Generative AI on Creativity in Software Development: A Research Agenda}.
\newblock  (\bibinfo{year}{2025}).
\newblock


\bibitem[Jimenez et~al\mbox{.}(2023)]%
        {swebench}
\bibfield{author}{\bibinfo{person}{Carlos~E Jimenez}, \bibinfo{person}{John Yang}, \bibinfo{person}{Alexander Wettig}, \bibinfo{person}{Shunyu Yao}, \bibinfo{person}{Kexin Pei}, \bibinfo{person}{Ofir Press}, {and} \bibinfo{person}{Karthik Narasimhan}.} \bibinfo{year}{2023}\natexlab{}.
\newblock \showarticletitle{Swe-bench: Can language models resolve real-world github issues?}
\newblock \bibinfo{journal}{\emph{arXiv preprint arXiv:2310.06770}} (\bibinfo{year}{2023}).
\newblock


\bibitem[Just et~al\mbox{.}(2014)]%
        {defects4j}
\bibfield{author}{\bibinfo{person}{Ren{\'e} Just}, \bibinfo{person}{Darioush Jalali}, {and} \bibinfo{person}{Michael~D Ernst}.} \bibinfo{year}{2014}\natexlab{}.
\newblock \showarticletitle{Defects4J: A database of existing faults to enable controlled testing studies for Java programs}. In \bibinfo{booktitle}{\emph{Proceedings of the 2014 international symposium on software testing and analysis}}. \bibinfo{pages}{437--440}.
\newblock


\bibitem[Le~Goues et~al\mbox{.}(2019)]%
        {le2019automated}
\bibfield{author}{\bibinfo{person}{Claire Le~Goues}, \bibinfo{person}{Michael Pradel}, {and} \bibinfo{person}{Abhik Roychoudhury}.} \bibinfo{year}{2019}\natexlab{}.
\newblock \showarticletitle{Automated program repair}.
\newblock \bibinfo{journal}{\emph{Commun. ACM}} \bibinfo{volume}{62}, \bibinfo{number}{12} (\bibinfo{year}{2019}), \bibinfo{pages}{56--65}.
\newblock


\bibitem[Li et~al\mbox{.}(2024)]%
        {cwebenchjava}
\bibfield{author}{\bibinfo{person}{Ziyang Li}, \bibinfo{person}{Saikat Dutta}, {and} \bibinfo{person}{Mayur Naik}.} \bibinfo{year}{2024}\natexlab{}.
\newblock \showarticletitle{IRIS: LLM-assisted static analysis for detecting security vulnerabilities}.
\newblock \bibinfo{journal}{\emph{arXiv preprint arXiv:2405.17238}} (\bibinfo{year}{2024}).
\newblock


\bibitem[Li et~al\mbox{.}(2021)]%
        {sysevr}
\bibfield{author}{\bibinfo{person}{Zhen Li}, \bibinfo{person}{Deqing Zou}, \bibinfo{person}{Shouhuai Xu}, \bibinfo{person}{Hai Jin}, \bibinfo{person}{Yawei Zhu}, {and} \bibinfo{person}{Zhaoxuan Chen}.} \bibinfo{year}{2021}\natexlab{}.
\newblock \showarticletitle{Sysevr: A framework for using deep learning to detect software vulnerabilities}.
\newblock \bibinfo{journal}{\emph{IEEE Transactions on Dependable and Secure Computing}} \bibinfo{volume}{19}, \bibinfo{number}{4} (\bibinfo{year}{2021}), \bibinfo{pages}{2244--2258}.
\newblock


\bibitem[Lin et~al\mbox{.}(2017)]%
        {quixbugs}
\bibfield{author}{\bibinfo{person}{Derrick Lin}, \bibinfo{person}{James Koppel}, \bibinfo{person}{Angela Chen}, {and} \bibinfo{person}{Armando Solar-Lezama}.} \bibinfo{year}{2017}\natexlab{}.
\newblock \showarticletitle{QuixBugs: A multi-lingual program repair benchmark set based on the Quixey Challenge}. In \bibinfo{booktitle}{\emph{Proceedings Companion of the 2017 ACM SIGPLAN international conference on systems, programming, languages, and applications: software for humanity}}. \bibinfo{pages}{55--56}.
\newblock


\bibitem[Lin et~al\mbox{.}(2025)]%
        {seagent}
\bibfield{author}{\bibinfo{person}{Jiaye Lin}, \bibinfo{person}{Yifu Guo}, \bibinfo{person}{Yuzhen Han}, \bibinfo{person}{Sen Hu}, \bibinfo{person}{Ziyi Ni}, \bibinfo{person}{Licheng Wang}, \bibinfo{person}{Mingguang Chen}, \bibinfo{person}{Hongzhang Liu}, \bibinfo{person}{Ronghao Chen}, \bibinfo{person}{Yangfan He}, \bibinfo{person}{Daxin Jiang}, \bibinfo{person}{Binxing Jiao}, \bibinfo{person}{Chen Hu}, {and} \bibinfo{person}{Huacan Wang}.} \bibinfo{year}{2025}\natexlab{}.
\newblock \bibinfo{title}{SE-Agent: Self-Evolution Trajectory Optimization in Multi-Step Reasoning with LLM-Based Agents}.
\newblock
\showeprint[arxiv]{2508.02085}~[cs.AI]
\urldef\tempurl%
\url{https://arxiv.org/abs/2508.02085}
\showURL{%
\tempurl}


\bibitem[Liu et~al\mbox{.}(2024)]%
        {liu2024large}
\bibfield{author}{\bibinfo{person}{Junwei Liu}, \bibinfo{person}{Kaixin Wang}, \bibinfo{person}{Yixuan Chen}, \bibinfo{person}{Xin Peng}, \bibinfo{person}{Zhenpeng Chen}, \bibinfo{person}{Lingming Zhang}, {and} \bibinfo{person}{Yiling Lou}.} \bibinfo{year}{2024}\natexlab{}.
\newblock \showarticletitle{Large language model-based agents for software engineering: A survey}.
\newblock \bibinfo{journal}{\emph{arXiv preprint arXiv:2409.02977}} (\bibinfo{year}{2024}).
\newblock


\bibitem[Liu et~al\mbox{.}(2019)]%
        {liu2019tbar}
\bibfield{author}{\bibinfo{person}{Kui Liu}, \bibinfo{person}{Anil Koyuncu}, \bibinfo{person}{Dongsun Kim}, {and} \bibinfo{person}{Tegawend{\'e}~F Bissyand{\'e}}.} \bibinfo{year}{2019}\natexlab{}.
\newblock \showarticletitle{TBar: Revisiting template-based automated program repair}. In \bibinfo{booktitle}{\emph{Proceedings of the 28th ACM SIGSOFT international symposium on software testing and analysis}}. \bibinfo{pages}{31--42}.
\newblock


\bibitem[Luo et~al\mbox{.}(2025)]%
        {agentsurvey2025}
\bibfield{author}{\bibinfo{person}{J. Luo}, \bibinfo{person}{W. Zhang}, \bibinfo{person}{Y. Yuan}, {et~al\mbox{.}}} \bibinfo{year}{2025}\natexlab{}.
\newblock \showarticletitle{Large Language Model Agent: A Survey on Methodology, Applications and Challenges}.
\newblock \bibinfo{journal}{\emph{arXiv preprint arXiv:2503.21460}} (\bibinfo{year}{2025}).
\newblock


\bibitem[Meng et~al\mbox{.}(2025)]%
        {meng2025empiricalstudyllmbasedagents}
\bibfield{author}{\bibinfo{person}{Xiangxin Meng}, \bibinfo{person}{Zexiong Ma}, \bibinfo{person}{Pengfei Gao}, {and} \bibinfo{person}{Chao Peng}.} \bibinfo{year}{2025}\natexlab{}.
\newblock \bibinfo{title}{An Empirical Study on LLM-based Agents for Automated Bug Fixing}.
\newblock
\showeprint[arxiv]{2411.10213}~[cs.SE]
\urldef\tempurl%
\url{https://arxiv.org/abs/2411.10213}
\showURL{%
\tempurl}


\bibitem[Mu et~al\mbox{.}(2025)]%
        {experepair}
\bibfield{author}{\bibinfo{person}{Fangwen Mu}, \bibinfo{person}{Junjie Wang}, \bibinfo{person}{Lin Shi}, \bibinfo{person}{Song Wang}, \bibinfo{person}{Shoubin Li}, {and} \bibinfo{person}{Qing Wang}.} \bibinfo{year}{2025}\natexlab{}.
\newblock \showarticletitle{EXPEREPAIR: Dual-Memory Enhanced LLM-based Repository-Level Program Repair}.
\newblock \bibinfo{journal}{\emph{arXiv preprint arXiv:2506.10484}} (\bibinfo{year}{2025}).
\newblock


\bibitem[Nguyen et~al\mbox{.}(2025)]%
        {projectdev}
\bibfield{author}{\bibinfo{person}{Minh~Huynh Nguyen}, \bibinfo{person}{Thang~Phan Chau}, \bibinfo{person}{Phong~X Nguyen}, {and} \bibinfo{person}{Nghi~DQ Bui}.} \bibinfo{year}{2025}\natexlab{}.
\newblock \showarticletitle{Agilecoder: Dynamic collaborative agents for software development based on agile methodology}. In \bibinfo{booktitle}{\emph{2025 IEEE/ACM Second International Conference on AI Foundation Models and Software Engineering (Forge)}}. IEEE, \bibinfo{pages}{156--167}.
\newblock


\bibitem[Pabba et~al\mbox{.}(2025)]%
        {semagent}
\bibfield{author}{\bibinfo{person}{Anvith Pabba}, \bibinfo{person}{Alex Mathai}, \bibinfo{person}{Anindya Chakraborty}, {and} \bibinfo{person}{Baishakhi Ray}.} \bibinfo{year}{2025}\natexlab{}.
\newblock \bibinfo{title}{SemAgent: A Semantics Aware Program Repair Agent}.
\newblock
\showeprint[arxiv]{2506.16650}~[cs.SE]
\urldef\tempurl%
\url{https://arxiv.org/abs/2506.16650}
\showURL{%
\tempurl}


\bibitem[Qian et~al\mbox{.}(2023a)]%
        {srdd}
\bibfield{author}{\bibinfo{person}{Chen Qian}, \bibinfo{person}{Xin Cong}, \bibinfo{person}{Cheng Yang}, \bibinfo{person}{Weize Chen}, \bibinfo{person}{Yusheng Su}, \bibinfo{person}{Juyuan Xu}, \bibinfo{person}{Zhiyuan Liu}, {and} \bibinfo{person}{Maosong Sun}.} \bibinfo{year}{2023}\natexlab{a}.
\newblock \showarticletitle{Communicative agents for software development}.
\newblock \bibinfo{journal}{\emph{arXiv preprint arXiv:2307.07924}} \bibinfo{volume}{6}, \bibinfo{number}{3} (\bibinfo{year}{2023}), \bibinfo{pages}{1}.
\newblock


\bibitem[Qian et~al\mbox{.}(2023b)]%
        {chatdev}
\bibfield{author}{\bibinfo{person}{Cheng Qian}, \bibinfo{person}{Wei Liu}, \bibinfo{person}{Hongzhang Liu}, \bibinfo{person}{Nuo Chen}, \bibinfo{person}{Yufan Dang}, \bibinfo{person}{Jiahao Li}, \bibinfo{person}{Cheng Yang}, \bibinfo{person}{Weize Chen}, \bibinfo{person}{Yusheng Su}, \bibinfo{person}{Xin Cong}, \bibinfo{person}{Juyuan Xu}, \bibinfo{person}{Dahai Li}, \bibinfo{person}{Zhiyuan Liu}, {and} \bibinfo{person}{Maosong Sun}.} \bibinfo{year}{2023}\natexlab{b}.
\newblock \showarticletitle{ChatDev: Communicative Agents for Software Development}. In \bibinfo{booktitle}{\emph{Annual Meeting of the Association for Computational Linguistics}}.
\newblock
\urldef\tempurl%
\url{https://api.semanticscholar.org/CorpusID:270257715}
\showURL{%
\tempurl}


\bibitem[Srivastava et~al\mbox{.}(2023)]%
        {bigbench}
\bibfield{author}{\bibinfo{person}{Aarohi Srivastava}, \bibinfo{person}{Abhinav Rastogi}, \bibinfo{person}{Abhishek Rao}, \bibinfo{person}{Abu Awal~Md Shoeb}, \bibinfo{person}{Abubakar Abid}, \bibinfo{person}{Adam Fisch}, \bibinfo{person}{Adam~R Brown}, \bibinfo{person}{Adam Santoro}, \bibinfo{person}{Aditya Gupta}, \bibinfo{person}{Adri{\`a} Garriga-Alonso}, {et~al\mbox{.}}} \bibinfo{year}{2023}\natexlab{}.
\newblock \showarticletitle{Beyond the imitation game: Quantifying and extrapolating the capabilities of language models}.
\newblock \bibinfo{journal}{\emph{Transactions on machine learning research}} (\bibinfo{year}{2023}).
\newblock


\bibitem[Team et~al\mbox{.}(2025)]%
        {trae}
\bibfield{author}{\bibinfo{person}{Trae~Research Team}, \bibinfo{person}{Pengfei Gao}, \bibinfo{person}{Zhao Tian}, \bibinfo{person}{Xiangxin Meng}, \bibinfo{person}{Xinchen Wang}, \bibinfo{person}{Ruida Hu}, \bibinfo{person}{Yuanan Xiao}, \bibinfo{person}{Yizhou Liu}, \bibinfo{person}{Zhao Zhang}, \bibinfo{person}{Junjie Chen}, \bibinfo{person}{Cuiyun Gao}, \bibinfo{person}{Yun Lin}, \bibinfo{person}{Yingfei Xiong}, \bibinfo{person}{Chao Peng}, {and} \bibinfo{person}{Xia Liu}.} \bibinfo{year}{2025}\natexlab{}.
\newblock \showarticletitle{Trae Agent: An LLM-based Agent for Software Engineering with Test-time Scaling}.
\newblock  (\bibinfo{year}{2025}).
\newblock
\showeprint[arxiv]{2507.23370}~[cs.SE]
\urldef\tempurl%
\url{https://arxiv.org/abs/2507.23370}
\showURL{%
\tempurl}


\bibitem[Wang et~al\mbox{.}(2025)]%
        {openhands}
\bibfield{author}{\bibinfo{person}{Xingyao Wang}, \bibinfo{person}{Boxuan Li}, \bibinfo{person}{Yufan Song}, \bibinfo{person}{Frank~F. Xu}, \bibinfo{person}{Xiangru Tang}, \bibinfo{person}{Mingchen Zhuge}, \bibinfo{person}{Jiayi Pan}, \bibinfo{person}{Yueqi Song}, \bibinfo{person}{Bowen Li}, \bibinfo{person}{Jaskirat Singh}, \bibinfo{person}{Hoang~H. Tran}, \bibinfo{person}{Fuqiang Li}, \bibinfo{person}{Ren Ma}, \bibinfo{person}{Mingzhang Zheng}, \bibinfo{person}{Bill Qian}, \bibinfo{person}{Yanjun Shao}, \bibinfo{person}{Niklas Muennighoff}, \bibinfo{person}{Yizhe Zhang}, \bibinfo{person}{Binyuan Hui}, \bibinfo{person}{Junyang Lin}, \bibinfo{person}{Robert Brennan}, \bibinfo{person}{Hao Peng}, \bibinfo{person}{Heng Ji}, {and} \bibinfo{person}{Graham Neubig}.} \bibinfo{year}{2025}\natexlab{}.
\newblock \bibinfo{title}{OpenHands: An Open Platform for AI Software Developers as Generalist Agents}.
\newblock
\showeprint[arxiv]{2407.16741}~[cs.SE]
\urldef\tempurl%
\url{https://arxiv.org/abs/2407.16741}
\showURL{%
\tempurl}


\bibitem[Wei et~al\mbox{.}(2025)]%
        {llmsmartaudit}
\bibfield{author}{\bibinfo{person}{Zhiyuan Wei}, \bibinfo{person}{Jing Sun}, \bibinfo{person}{Yuqiang Sun}, \bibinfo{person}{Ye Liu}, \bibinfo{person}{Daoyuan Wu}, \bibinfo{person}{Zijian Zhang}, \bibinfo{person}{Xianhao Zhang}, \bibinfo{person}{Meng Li}, \bibinfo{person}{Yang Liu}, \bibinfo{person}{Chunmiao Li}, {et~al\mbox{.}}} \bibinfo{year}{2025}\natexlab{}.
\newblock \showarticletitle{Advanced smart contract vulnerability detection via llm-powered multi-agent systems}.
\newblock \bibinfo{journal}{\emph{IEEE Transactions on Software Engineering}} (\bibinfo{year}{2025}).
\newblock


\bibitem[Widyasari et~al\mbox{.}(2020)]%
        {bugsinpy}
\bibfield{author}{\bibinfo{person}{Ratnadira Widyasari}, \bibinfo{person}{Sheng~Qin Sim}, \bibinfo{person}{Camellia Lok}, \bibinfo{person}{Haodi Qi}, \bibinfo{person}{Jack Phan}, \bibinfo{person}{Qijin Tay}, \bibinfo{person}{Constance Tan}, \bibinfo{person}{Fiona Wee}, \bibinfo{person}{Jodie~Ethelda Tan}, \bibinfo{person}{Yuheng Yieh}, {et~al\mbox{.}}} \bibinfo{year}{2020}\natexlab{}.
\newblock \showarticletitle{Bugsinpy: a database of existing bugs in python programs to enable controlled testing and debugging studies}. In \bibinfo{booktitle}{\emph{Proceedings of the 28th ACM joint meeting on european software engineering conference and symposium on the foundations of software engineering}}. \bibinfo{pages}{1556--1560}.
\newblock


\bibitem[Xia et~al\mbox{.}(2024a)]%
        {swebenchlites}
\bibfield{author}{\bibinfo{person}{Chunqiu~Steven Xia}, \bibinfo{person}{Yinlin Deng}, \bibinfo{person}{Soren Dunn}, {and} \bibinfo{person}{Lingming Zhang}.} \bibinfo{year}{2024}\natexlab{a}.
\newblock \showarticletitle{Agentless: Demystifying llm-based software engineering agents}.
\newblock \bibinfo{journal}{\emph{arXiv preprint arXiv:2407.01489}} (\bibinfo{year}{2024}).
\newblock


\bibitem[Xia et~al\mbox{.}(2024b)]%
        {xia2024agentless}
\bibfield{author}{\bibinfo{person}{Chunqiu~Steven Xia}, \bibinfo{person}{Yinlin Deng}, \bibinfo{person}{Soren Dunn}, {and} \bibinfo{person}{Lingming Zhang}.} \bibinfo{year}{2024}\natexlab{b}.
\newblock \showarticletitle{Agentless: Demystifying llm-based software engineering agents}.
\newblock \bibinfo{journal}{\emph{arXiv preprint arXiv:2407.01489}} (\bibinfo{year}{2024}).
\newblock


\bibitem[Yang et~al\mbox{.}(2024)]%
        {sweagent}
\bibfield{author}{\bibinfo{person}{John Yang}, \bibinfo{person}{Carlos~E. Jimenez}, \bibinfo{person}{Alexander Wettig}, \bibinfo{person}{Kilian Lieret}, \bibinfo{person}{Shunyu Yao}, \bibinfo{person}{Karthik Narasimhan}, {and} \bibinfo{person}{Ofir Press}.} \bibinfo{year}{2024}\natexlab{}.
\newblock \bibinfo{title}{SWE-agent: Agent-Computer Interfaces Enable Automated Software Engineering}.
\newblock
\showeprint[arxiv]{2405.15793}~[cs.SE]
\urldef\tempurl%
\url{https://arxiv.org/abs/2405.15793}
\showURL{%
\tempurl}


\bibitem[Yu et~al\mbox{.}(2025)]%
        {orcaloca}
\bibfield{author}{\bibinfo{person}{Zhongming Yu}, \bibinfo{person}{Hejia Zhang}, \bibinfo{person}{Yujie Zhao}, \bibinfo{person}{Hanxian Huang}, \bibinfo{person}{Matrix Yao}, \bibinfo{person}{Ke Ding}, {and} \bibinfo{person}{Jishen Zhao}.} \bibinfo{year}{2025}\natexlab{}.
\newblock \bibinfo{title}{OrcaLoca: An LLM Agent Framework for Software Issue Localization}.
\newblock
\showeprint[arxiv]{2502.00350}~[cs.SE]
\urldef\tempurl%
\url{https://arxiv.org/abs/2502.00350}
\showURL{%
\tempurl}


\bibitem[Zhang et~al\mbox{.}(2023)]%
        {zhang2023survey}
\bibfield{author}{\bibinfo{person}{Quanjun Zhang}, \bibinfo{person}{Chunrong Fang}, \bibinfo{person}{Yuxiang Ma}, \bibinfo{person}{Weisong Sun}, {and} \bibinfo{person}{Zhenyu Chen}.} \bibinfo{year}{2023}\natexlab{}.
\newblock \showarticletitle{A survey of learning-based automated program repair}.
\newblock \bibinfo{journal}{\emph{ACM Transactions on Software Engineering and Methodology}} \bibinfo{volume}{33}, \bibinfo{number}{2} (\bibinfo{year}{2023}), \bibinfo{pages}{1--69}.
\newblock


\bibitem[Zhang et~al\mbox{.}(2024)]%
        {caasd}
\bibfield{author}{\bibinfo{person}{Simiao Zhang}, \bibinfo{person}{Jiaping Wang}, \bibinfo{person}{Guoliang Dong}, \bibinfo{person}{Jun Sun}, \bibinfo{person}{Yueling Zhang}, {and} \bibinfo{person}{Geguang Pu}.} \bibinfo{year}{2024}\natexlab{}.
\newblock \showarticletitle{Experimenting a new programming practice with llms}.
\newblock \bibinfo{journal}{\emph{arXiv preprint arXiv:2401.01062}} (\bibinfo{year}{2024}).
\newblock


\bibitem[Zhang et~al\mbox{.}(2025)]%
        {agentorchestra}
\bibfield{author}{\bibinfo{person}{Wentao Zhang}, \bibinfo{person}{Liang Zeng}, \bibinfo{person}{Yuzhen Xiao}, \bibinfo{person}{Yongcong Li}, \bibinfo{person}{Ce Cui}, \bibinfo{person}{Yilei Zhao}, \bibinfo{person}{Rui Hu}, \bibinfo{person}{Yang Liu}, \bibinfo{person}{Yahui Zhou}, {and} \bibinfo{person}{Bo An}.} \bibinfo{year}{2025}\natexlab{}.
\newblock \bibinfo{title}{AgentOrchestra: Orchestrating Hierarchical Multi-Agent Intelligence with the Tool-Environment-Agent(TEA) Protocol}.
\newblock
\showeprint[arxiv]{2506.12508}~[cs.AI]
\urldef\tempurl%
\url{https://arxiv.org/abs/2506.12508}
\showURL{%
\tempurl}


\bibitem[Zhuge et~al\mbox{.}({[n.\,d.]})]%
        {gptswarm}
\bibfield{author}{\bibinfo{person}{Mingchen Zhuge}, \bibinfo{person}{Wenyi Wang}, \bibinfo{person}{Louis Kirsch}, \bibinfo{person}{Francesco Faccio}, \bibinfo{person}{Dmitrii Khizbullin}, {and} \bibinfo{person}{J{\"u}rgen Schmidhuber}.} \bibinfo{year}{[n.\,d.]}\natexlab{}.
\newblock \showarticletitle{GPTSwarm: Language Agents as Optimizable Graphs}. In \bibinfo{booktitle}{\emph{Forty-first International Conference on Machine Learning}}.
\newblock


\end{thebibliography}

\end{document}